\documentclass[11pt]{article}
\usepackage[IL2]{fontenc}
\usepackage{amsmath, amssymb, amsthm, amsfonts}
\usepackage{bbm}
\usepackage{amscd}
\usepackage{mathrsfs}
\usepackage{float}

\usepackage{comment} 
\usepackage{ifthen}
\usepackage{tikz}
\usetikzlibrary{positioning,decorations.pathreplacing}
\usepackage{ bbold }
\usepackage{appendix}
\usepackage{graphicx}
\usepackage{color}
\usepackage{epstopdf}
\usepackage{wrapfig}
\usepackage{paralist}
\usepackage{wasysym}
\usepackage[textsize=tiny]{todonotes}

\usepackage{listings} 

\usepackage{pdflscape} 

\usepackage{framed}
\usepackage[framemethod=tikz]{mdframed}
\usepackage[bottom]{footmisc}
\usepackage{enumitem}
\setitemize{noitemsep,topsep=3pt,parsep=3pt,partopsep=3pt}
\usepackage[font=small]{caption}
\usepackage{xspace}

\usepackage{thmtools} 
\usepackage{thm-restate} 

\definecolor{darkgreen}{rgb}{0,0.5,0}
\definecolor{darkblue}{rgb}{0,0,0.6}
\usepackage{hyperref}
\hypersetup{
    unicode=false,          
    colorlinks=true,        
    linkcolor=darkblue,          
    citecolor=darkgreen,        
    filecolor=magenta,      
    urlcolor=cyan           
}

\newtheorem{theorem}{Theorem}[section]
\newtheorem{lemma}[theorem]{Lemma}
\newtheorem{meta-theorem}[theorem]{Meta-Theorem}

\newtheorem{observation}[theorem]{Observation}
\newtheorem{definition}[theorem]{Definition}

\usepackage[capitalize, nameinlink,noabbrev]{cleveref}

\crefname{theorem}{Theorem}{Theorems}
\crefname{proposition}{Proposition}{Propositions}
\crefname{observation}{Observation}{Observations}
\crefname{lemma}{Lemma}{Lemmas}
\crefname{claim}{Claim}{Claims}
\crefname{problem}{Problem}{Problems}
\crefname{conjecture}{Conjecture}{Conjectures}
\crefname{question}{Question}{Questions}
\crefname{example}{Example}{Examples}
\crefname{fact}{Fact}{Facts}

\definecolor{darkgreen}{rgb}{0,0.5,0}

\usepackage{algcompatible}
\algnewcommand\algorithmicswitch{\textbf{switch}}
\algnewcommand\algorithmiccase{\textbf{case}}

\algdef{SE}[SWITCH]{Switch}{EndSwitch}[1]{\algorithmicswitch\ #1\ \algorithmicdo}{\algorithmicend\ \algorithmicswitch}%
\algdef{SE}[CASE]{Case}{EndCase}[1]{\algorithmiccase\ #1}{\algorithmicend\ \algorithmiccase}%
\algtext*{EndSwitch}%
\algtext*{EndCase}%

\newcommand{\eps}{\varepsilon}

\newcommand{\poly}{\operatorname{poly}}

\renewcommand{\phi}{\varphi}

\renewcommand{\paragraph}[1]{\bigskip \noindent {\bf #1}:}

\newcommand{\FullOrShort}{full}
\ifthenelse{\equal{\FullOrShort}{full}}{
  
  \newcommand{\fullOnly}[1]{#1}
  \newcommand{\shortOnly}[1]{}

  }{

    \newcommand{\fullOnly}[1]{}
    \newcommand{\IncludePictures}[1]{}
   
  }

\newcommand{\Pot}

\usepackage{algorithm}
\usepackage[noend]{algpseudocode}

\usepackage[letterpaper,margin=1.00in]{geometry}

\date{ }

\newcommand{\strtch}{$O(\log n \cdot \text{poly}(\log \log n))$}

\begin{document}
\date{}
\title{Parallel Batch-Dynamic Algorithms for Spanners, and Extensions}
\author{Mohsen Ghaffari \\ \small MIT \\ \small ghaffari@mit.edu\and Jaehyun Koo \\ \small MIT \\ \small koosaga@mit.edu}
\maketitle

\begin{abstract} 
This paper presents the first parallel batch-dynamic algorithms for computing spanners and sparsifiers. Our algorithms process any batch of edge insertions and deletions in an $n$-node undirected graph, in $\poly(\log n)$ depth and using amortized work near-linear in the batch size. Our concrete results are as follows:

\begin{itemize}
\item Our base algorithm maintains a spanner with $(2k-1)$ stretch and $\tilde{O}(n^{1+1/k})$ edges, for any $k\geq 1$.
\item Our first extension maintains a sparse spanner with only $O(n)$ edges, and $\tilde{O}(\log n)$ stretch.
\item Our second extension maintains a $t$-bundle of spanners ---i.e., $t$ spanners, each of which is the spanner of the graph remaining after removing the previous ones---and allows us to maintain cut/spectral sparsifiers with $\tilde{O}(n)$ edges. 
\end{itemize}
\end{abstract}

\maketitle

\section{Introduction}
Graph sparsifications that approximately preserve certain graph properties have been among the main graph algorithm tools developed and used over the past two to three decades. The leading notions have been \textit{spanners}, which reduce the number of edges while approximately keeping pairwise distances, and \textit{cut sparsifiers} (sometimes simply called sparsifiers, and their strengthening spectral sparsifiers), which reduce the number of edges while approximately preserving the size of all cuts. Contributing to the burgeoning field of \textit{batch-dynamic parallel algorithms}---see e.g. \cite{acar2019parallel,acar2020parallel,dhulipala2021parallel,tseng2022parallel,liu2022parallel,ghaffari2023nearly, anderson2024deterministic, ghaffari2024parallel} for some recent work---we present the first such algorithms for computing/maintaining spanners and sparsifiers. Our algorithms process each batch of edge insertions and deletions in polylogarithmic depth, using amortized work bounds linear in the batch size, up to logarithmic factors. Hence, they provide batch-dynamic parallel algorithms for these problems with near-optimal amortized runtime, for any number of processors. We review the context and state of the art, and then formally state our results.

\subsection{Context: Problems and the Model}

\noindent\textbf{Problem Definitions---Spanners}: Given an undirected unweighted graph $G=(V, E)$, a subgraph $H=(V, E')$ of it is called a $k$-spanner, or sometimes a spanner with stretch $k$, if for all $u, v\in V$, it holds that $dist_H(u,v) \leq k \cdot dist_G(u,v)$. Here, $dist_H$ is the distance metric in graph $H$. The size of the spanner is simply the number of its edges $|E'|$. It is known that every $n$-node graph has a $(2k-1)$-spanner of size $O(n^{1+1/k})$\cite{althofer1993sparse}, and this size bound is tight conditioned on a girth conjecture of Erd\H{o}s~\cite{erdos1963extremal}. Of particular interest is the case of spanners with $O(n)$ edges, which are often called sparse spanners, and their existence follows by setting $k = \Theta({\log n})$. Spanners have found numerous applications, including packet routing, network synchronizers, and algorithms for various graph problems. See, e.g.,~\cite{AwerbuchBCP93,AwerbuchP92,cohen93,thorup2005approximate,ThorupZ01routing,GavoillePPR04,LenzenL18} and the recent survey of Ahmed et al.~\cite{ahmed2020graph}. 

\paragraph{Problem Definitions---Sparsifiers} Given an undirected unweighted graph $G=(V, E)$, a weighted subgraph $H=(V, E', w)$ of it is called a cut sparsifier if for every nonempty set $S\subseteq V$, we have $(1-\eps) cut_{G}(S, V\setminus S) \leq cut_{H}(S, V\setminus S) \leq (1+\eps) cut_{G}(S, V\setminus S)$. Here, $\eps>0$ is an arbitrarily small parameter usually regarded as a constant or inverse-polylogarithmic, $cut_{G}(S, V\setminus S)$ and $cut_{H}(S, V\setminus S)$ denote the total number of edges in $G$, and the total weight of edges in $H$ respectively, with exactly one endpoint in $S$. Sometimes this is called a $(1\pm \eps)$-sparsifier to make the $\eps$ parameter explicit. The size of the sparsifier $H$ is simply its number of edges $|E'|$. Results of Benczur and Karger~\cite{benczur1996approximating} showed that every $n$-node graph has a sparsifier with size $O(n\log n/\eps^2)$, and the bound was sharpened later to $O(n/\eps^2)$\cite{batson2009twice} (even for the stronger notion of spectral sparsifiers; see \Cref{sec:bundlesANDsparsifier} for a definition review).

\paragraph{Batch-Dynamic Parallel Algorithms}  We seek parallel algorithms that maintain such spanners and sparsifiers as the graph undergoes updates. For parallel computation, we follow the standard work-depth terminology~\cite{blelloch1996programming}. An algorithm's work $W$ is its total number of operations, and its \textit{depth} $D$ is the length of the longest chain of operations with sequential dependencies. The time to run the algorithm on a setup with $p$ processors is at least $\max\{W(\mathcal{A})/p, D(\mathcal{A})\}$, and at most $W(\mathcal{A})/p + D(\mathcal{A})\leq 2 \max\{W(\mathcal{A})/p, D(\mathcal{A})\}$~\cite{brent1974parallel}.

We want dynamic algorithms that quickly adjust the solution to graph updates---i.e., potentially large batches of edge insertions and deletions---by leveraging parallelism as much as possible. Any batch of $b$ updates necessitates at least $b$ work. We give algorithms that process each such batch using $b\cdot \poly(\log n)$ amortized work and in $\poly(\log n)$ depth. These algorithms would run in $\tilde{O}(b/p)$ amortized time, given $p$ processors. This is a nearly optimal amortized runtime for any number of given processors, up to $\log n$ factors.

\subsection{State of the Art and Related Work}
Batch-dynamic parallel algorithms strengthen two seemingly unrelated algorithms: \textit{static parallel algorithms}, which receive and process the entire graph once, and \textit{dynamic sequential algorithms}, which receive and process the edge insertion/deletion updates one by one. So, we review each area's state of the art before presenting our results.

\paragraph{Static Parallel Algorithms---Spanners} A classic randomized algorithm of Baswana and Sen~\cite{baswana2007simple} constructs a $(2k-1)$-spanner with size $O(k \cdot n^{1+1/k})$, in near-ideal depth and work bounds of $\poly(\log n)$ and $\tilde{O}(m)$, with high probability. Miller et al.~\cite{miller2015improved} improved the size to $O(n^{1+1/k})$ at the expense of increasing stretch to $O(k)$, while keeping the depth and work bounds near-ideal as before. Elkin and Neiman~\cite{10.1145/3274651} adjusted this construction to output a $(2k-1)$ spanner with size $O(n^{1+1/k}/\eps)$ with probability $1-\eps$, again in the same depth and work bound.\footnote{Some of these algorithms are described originally in a distributed model, with $O(k)$ rounds of computation. Still, it is easy to see that they can be implemented in $\poly(\log n)$ depth and $\tilde{O}(m)$ work in any PRAM model.} In particular, this can produce a sparse spanner with the optimal stretch. 

Of notable relevance for our case is that in the works of Miller et al.~\cite{miller2015improved} and Elkin and Neiman~\cite{10.1145/3274651}, there is an essentially black-box reduction from spanner computation to single-source shortest-path tree computation in a related unweighted graph (with exponentially-distributed head starts in the style of \cite{miller2013parallel}), up to a $\poly(\log n)$ distance. We make use of this approach in our base batch-dynamic parallel spanner algorithm. 

\paragraph{Static Parallel Algorithms---Sparsifiers} Koutis~\cite{koutis2014simple} showed an algorithm that computes a $(1\pm \eps)$ spectral/cut sparsifier with $O(n\poly(\log n)/\eps^2)$ edges, for any constant $\eps>0$, using the near-ideal depth and work bounds of $\poly(\log n)$ and $\tilde{O}(m)$. The basic ingredient is the iterative procedure where one repeatedly computes a spanner and removes it for $\poly(\log n, 1/\eps)$ iterations. These sets of computed spanners are called spanner bundles. Following this outline, to provide a batch-dynamic parallel sparsifier, our core task will be to give a batch-dynamic parallel algorithm for computing spanner bundles. This does not trivially follow from the batch-dynamic parallel spanner, since propagating the updates through the iterations of spanner packing can blow up the number of updates.


\paragraph{Dynamic Sequential Algorithms---Spanners} Baswana~\cite{baswana2006dynamic} gave the first dynamic sequential algorithm for $(2k-1)$-spanners of expected size $O(k \cdot n^{1+1/k})$,  though limited to only decremental updates, in expected $\poly(\log n)$ time per edge deletion update. Elkin~\cite{elkin2011streaming} gave the dynamic sequential algorithm for $(2k-1)$-spanner of size $O((k\log n)^{1-1/k} \cdot n^{1+1/k})$, with $O(1)$ expected time per edge insertion, but the update time for edge deletion is rather high and we ignore that here. Baswana et al.\cite{baswana2008fully, baswana2012fully} gave the first fully-dynamic sequential algorithm, which processes each insertion or deletion in $\poly(\log n)$ amortized time and maintains a $(2k-1)$-spanner with size $O(k\log n \cdot n^{1+1/k})$. They apply a classical idea of Bentley and Saxe \cite{bentley1980decomposable} that reduces the fully-dynamic amortized case to the incremental-only, which we also employ in our algorithms. Forster and Goranci \cite{forster2019dynamic} improved on this and gave a fully dynamic algorithm with $\poly(\log n)$-time amortized to keep a spanner of size $O(\log n \cdot n^{1+1/k})$. Their dynamic algorithm leverages the reduction from spanners to decremental single-source shortest-path trees mentioned above (from the static algorithms of \cite{miller2015improved, 10.1145/3274651}), where the decremental shortest-path tree part is maintained with a classic approach of Even and Shiloach~\cite{shiloach1981line}. We also use this idea, though we must make it work in the batch-dynamic setting with simultaneous updates.

Some other related works that should be mentioned here are: (I) sequential dynamic spanner algorithms with worst-case update bounds~\cite{bernstein2021deamortization,bodwin2016fully}, and (II) sequential dynamic spanner algorithms against an adaptive adversary that can adapt its updates to the results produced (and thus the randomness used) by the randomized dynamic algorithm~\cite{bhattacharya2022simple,bernstein2020fully}.



\paragraph{Dynamic Sequential Algorithms---Sparsifiers} A work of Abraham et al.~\cite{abraham2016fully} showed how to maintain $(1\pm \eps)$ spectral/cut sparsifier of size $O(n\poly(\log n)/\eps^2)$ in a fully dynamic setting, and with $\poly(\log n)$ amortized work per edge insertion/deletion, by making dynamic the aforementioned iterative spanner packing approach of Koutis~\cite{koutis2014simple} dynamic. In particular, this involved creating a dynamic algorithm for computing spanner bundles. Following this outline, similarly, the core task for our batch-dynamic parallel sparsifier construction will be to have a batch-dynamic parallel algorithm for spanner bundles.

\subsection{Our Results}
Below, we state our results for spanners and related problems. The proofs of these results are provided in Part III of the thesis (Sections 7 to 10).

All of our algorithms in this part work with $\poly(\log n)$-depth per batch and using amortized work $b\cdot \poly(\log n)$ for each batch, where $b$ denotes the size of that batch. We present our results in three parts: (A) base spanner algorithm, (B) the first extension, sparse and ultra-sparse spanners, (C) the second extension, spanner bundles and sparsifiers.

\paragraph{(A) Base Spanner Algorithm} Our base result is a batch-dynamic parallel algorithm for spanners, under edge insertion and deletion batches, with the near-optimal stretch-size tradeoff:

\begin{restatable}[\textbf{Parallel Batch-Dynamic Spanners}]{theorem}{mainnear}
\label{thm:mainnear}
    There is a randomized parallel batch-dynamic data structure which, given an unweighted graph $G = (V, E)$ where $|V| = n, |E| = m$, maintains a $(2k-1)$-spanner with $O(n^{1 + 1/k} \log n)$ expected number of edges. Specifically, the algorithm supports the following interfaces:

    \begin{itemize}
        \item After the initialization, the algorithm returns a set of edges forming a $(2k-1)$-spanner of the given graph, which has $O(n^{1+1/k})$ expected number of edges.
        \item After any batch of edge updates, the algorithm returns a pair of edge sets $(\delta H_{ins}, \delta H_{del})$, which represent the set of edges that are newly inserted or deleted into the $(2k-1)$-spanner. The amortized size of $|\delta H_{ins}| + |\delta H_{del}|$ is at most $O(k \log^2 n)$ per edge in expectation.
    \end{itemize}

    The algorithm takes:
    \begin{itemize}
    \item for initialization, $O(m \log n)$ expected work, and $O(\log^2 n)$ worst-case depth,
    \item for any batch of edge updates, $O(k \log^2 n)$ expected amortized work per edge, and $O(k \log^2 n)$ worst-case depth for the entire batch.
    \end{itemize}
The above statements hold with high probability against an oblivious adversary.
\end{restatable}

As a key ingredient in the above batch-dynamic spanner algorithm, we provide a batch-parallel decremental algorithm for single-source shortest path tree under batches of edge deletions: \begin{restatable}[\textbf{Parallel Batch-Dynamic BFS}]{theorem}{estree}
\label{thm:estree}
    There is a deterministic parallel batch-dynamic decremental data structure which, given an unweighted directed graph $G = (V, E)$ where $|V| = n, |E| = m$, a source vertex $s \in V$, and a parameter $L \leq n$, maintains a shortest path tree of depth $L$.
    The algorithm takes:
\begin{itemize}
    \item for initialization, $O(m \log n)$ work, and $O(L \log n + \log^2 n)$ worst-case depth,
    \item for any batch of edge deletions, $O(L \log n)$ amortized work per deleted edge, and $O(L \log^2 n)$ worst-case depth for the entire batch.
\end{itemize}

\end{restatable}
\Cref{thm:estree} is in essence a batch-dynamic parallel variant of the classic sequential decremental-dynamic algorithm of Even and Shiloach~\cite{shiloach1981line}. The bounds in this result are particularly good when the maximum distance of interest $L$ is small, which is the case for our application.

\paragraph{(B) Extension to Sparse and Ultra-Sparse Spanners via Iterated Contractions} Our first extension allows us to reduce the number of edges to $O(n)$, with a near-optimal stretch of $\tilde{O}(\log n)$. Such spanners are often called sparse spanners and they are particularly useful in many applications --- in a sense, they have asymptotically the same number of edges as a spanning tree, while having nice distance properties. The algorithm is built via an iterative approach that reduces the number of vertices in the graph to $n/\poly(\log n)$ by repeatedly performing certain local clusterings and contractions, and then applies our base spanner algorithm on this smaller graph. The technical challenge is controlling how the updates propagate through this iterative scheme and avoiding their blow up (which would happen in any naive iterative scheme). Our second extension shows that, by adding one more layer of clustering and contractions, we can reduce the number of edges to $n + o(n)$, only at a slight expense of stretch and computation cost. Such spanners are often called ultra-sparse spanners, which provides an even stronger notion of sparsity in a way that almost all of its edges are a part of the spanning tree. Ultra-sparse spanners have emerged as a powerful tool for optimizing numerous fundamental graph problems, such as maximum flow, min-cut problems, and approximate shortest path problems \cite{daitch2008faster, li2020faster, peng2016approximate}.

\begin{restatable}[\textbf{Parallel Batch-Dynamic Sparse Spanners}]{theorem}{maintrue}\label{thm:maintrue}
    There is a randomized parallel batch-dynamic data structure which, given an unweighted graph $G = (V, E)$ where $|V| = n, |E| = m$, maintains a \strtch-spanner of $O(n)$ expected number of edges. Specifically, the algorithm supports the following interfaces:

    \begin{itemize}
        \item After the initialization, the algorithm returns a set of edges forming a \strtch-spanner of the given graph, which has $O(n)$ expected number of edges.
        \item After any batch of edge updates, the algorithm returns a pair of edge sets $(\delta H_{ins}, \delta H_{del})$, representing the set of edges that are newly inserted or deleted into the \strtch-spanner. The amortized size of $|\delta H_{ins}| + |\delta H_{del}|$ is at most $O(\log^3 n)$ per edge in expectation.
    \end{itemize}

    The algorithm takes:
    \begin{itemize}
    \item for initialization, $O(m \log n)$ expected work, and $O(\log^2 n)$ worst-case depth,
    \item for any batch of edge updates, $O(\log^3 n \cdot \text{poly}(\log \log n))$ expected amortized work per updated edge, and $O(\log^3 n)$ worst-case depth for the entire batch,
    \end{itemize}
    The above statements hold with high probability against an oblivious adversary.
\end{restatable}

\begin{restatable}[\textbf{Parallel Batch-Dynamic Ultra-Sparse Spanners}]{theorem}{mainultra}\label{thm:mainultra}
    There is a randomized parallel batch-dynamic data structure which, given an unweighted graph $G = (V, E)$ where $|V| = n, |E| = m$ and an integer $2 \le x \le O(\frac{\log \log n}{(\log \log \log n)^2})$, maintains a \strtch-spanner of at most $n + O(n / x)$ edges. Specifically, the algorithm supports the following interfaces:

    \begin{itemize}
        \item After the initialization, the algorithm returns a set of edges forming a \strtch-spanner of the given graph, which has $n + O(n / x)$ expected number of edges.
        \item After any batch of edge updates, the algorithm returns a pair of edge sets $(\delta H_{ins}, \delta H_{del})$, representing the set of edges that are newly inserted or deleted into the \strtch-spanner. The amortized size of $|\delta H_{ins}| + |\delta H_{del}|$ is at most $O(\log^3 n \cdot 2^{O(x \log^2 x)})$ per edge in expectation.
    \end{itemize}

    The algorithm takes:
    \begin{itemize}
    \item for initialization, $O(m \log n)$ expected work, and $O(\log^2 n)$ worst-case depth,
    \item for any batch of edge updates, $O(\log^3 n \cdot \text{poly}(\log \log n) \cdot 2^{O(x \log^2 x)})$ expected amortized work per updated edge, and $O(\log^3 n)$ worst-case depth for the entire batch,
    \end{itemize}
    The above statements hold with high probability against an oblivious adversary.
\end{restatable}

\paragraph{(C) Extension to Spanner Bundles and Sparsifiers} Our second extension allows us to compute $t$-bundle spanners, i.e., a collection of $t$ spanners $H_1$, $H_2$, \dots, $H_t$ where each $H_i$ is a spanner of the graph $G\setminus (\cup_{j=1}^{i-1} H_j)$. As a result, and with some extra work, this allows us to provide a batch-dynamic parallel algorithm for spectral/cut sparsifiers, with near-optimal size.

\begin{restatable}[\textbf{Parallel Batch-Dynamic Spanner Bundles}]{theorem}{decrbundle}\label{thm:decr-bundle2}
        There is a parallel batch-dynamic decremental data structure which, given an unweighted graph $G = (V, E)$ where $|V| = n, |E| = m$, maintains a $t$-bundle spanner of at most $O(nt \log^3 n)$ expected number of edges. Specifically, the algorithm supports the following interfaces:

    \begin{itemize}
        \item After the initialization, the algorithm returns a set of edges forming a $t$-bundle spanner of the given graph, which has $O(nt \log n)$ edges.
        \item After each edge deletion updates, the algorithm returns a pair of edge sets $(\delta H_{ins}, \delta H_{del})$, representing the set of edges that are newly inserted or deleted into the $t$-bundle spanner. The amortized size of $|\delta H_{ins}| + |\delta H_{del}|$ is at most $O(1)$ per edge.
    \end{itemize}

    The algorithm takes:
\begin{itemize}
    \item for initialization, $O(m t \log^2 n)$ work, and $O(t \log^2 n)$ worst-case depth,
    \item for any batch of edge deletions, $O(t \log^3 n)$ expected amortized work per deleted edge, and $O(t \log^3 n)$ worst-case depth for the entire batch.
\end{itemize}
    The above statements hold with high probability against an oblivious adversary.
\end{restatable}

\begin{restatable}[\textbf{Parallel Batch-Dynamic Spectral Sparsifiers}]{theorem}{spectral}\label{thm:spectral}
        There is a parallel batch-dynamic data structure which, given an unweighted graph $G = (V, E)$ where $|V| = n, |E| = m$, maintains a $(1 \pm \epsilon)$ spectral sparsifier of at most $O(n \epsilon^{-2} \log^4 m \log^6 n)$ expected number of edges. Specifically, the algorithm supports the following interfaces:

    \begin{itemize}
        \item After the initialization, the algorithm returns a set of edges forming a $(1 \pm \epsilon)$-spectral sparsifier of the given graph, which has $O(n \epsilon^{-2} \log^3 m \log^4 n)$ expected number of edges.
        \item After any batch of edge updates, the algorithm returns a pair of edge sets $(\delta H_{ins}, \delta H_{del})$, representing the set of edges that are newly inserted or deleted into the $(1 \pm \epsilon)$-spectral sparsifier. The amortized size of $|\delta H_{ins}| + |\delta H_{del}|$ is at most $O(\log m)$ per edge.
    \end{itemize}

    The algorithm takes:
\begin{itemize}
    \item for initialization, $O(m \epsilon^{-2} \log^3 m \log^5 n)$ expected work, and $O(\epsilon^{-2} \log^3 m \log^5 n)$ worst-case depth,
    \item for any batch of edge updates, $O(\epsilon^{-2} \log^4 m \log^6 n)$ expected amortized work per updated edge, and $O(\epsilon^{-2} \log^3 m \log^6 n)$ worst-case depth for the entire batch,
\end{itemize}
    The above statements hold with high probability against an oblivious adversary.
\end{restatable}

\section{Preliminaries} \label{sec:prelim}
\paragraph{Binary Search Tree} We use the parallel red-black tree in \cite{PARK2001415} to deterministically maintain an ordered list. In CRCW PRAM, the algorithm takes $O(\log n)$ work per element and $O(\log n)$ depth in each batch operation. This result also implies an $O(\log n)$-depth and an $O(n \log n)$ work parallel sorting algorithm.

\paragraph{Hash Tables} We use the parallel randomized hash table in \cite{gil1991towards} for maintaining a dictionary. In CRCW PRAM, the algorithm takes $O(1)$ work per element and $O(\log^* n)$ depth in each batch operation. The work and depth bound hold with high probability.

\section{Spanners}
Here, we present our base batch-dynamic parallel spanner algorithm, thus proving \Cref{thm:mainnear}.

\subsection{Data Structures}
The following data structure maintains the list of in-edges from each vertex. The proof is deferred to the full version.

\begin{restatable}{lemma}{AppAdjList}
\label{lem:app_adjlist}
    There exists a deterministic parallel data structure that supports the following operations:
    \begin{itemize}
    \item \textsc{Initialize}($\{(v_1, p_1), (v_2, p_2), \ldots, (v_l, p_l)\}$): Initialize the data structure with $l$ elements $v_1, \ldots, v_l$, where each element is associated with corresponding priority $p_1, \ldots, p_l$. $p_i$ should be distinct and should be in range $[1, poly(n)]$. 
    \item \textsc{UpdateValue}$(k, v)$: Set the value of the element with $k$-th largest priority to $v$. 
    \item \textsc{UpdatePriority}$(k, p)$: Set the priority of the element with $k$-th largest priority to $p$. All priorities should remain distinct.
    \item \textsc{Query}$(k)$: Return the element with $k$-th largest priority.
    \item \textsc{Find}$(p)$: Return the element with priority $p$, along with the number of element with priority at least $p$.
    \item \textsc{NextWith}$(k, f)$: Given the position $k$ and a function $f$ that can be evaluated in $O(1)$ work, find the smallest $p \geq k$ such that $f(\textsc{Query}(p))= \textsf{TRUE}$. If such an element does not exist, return $p = l + 1$.
    \end{itemize}
    Each operation takes the following depth and work:
    \begin{itemize}
    \item \textsc{Initialize}($\{(v_1, p_1), (v_2, p_2), \ldots, (v_l, p_l)\}$) takes $O(l \log n)$ work with $O(\log n)$ depth.
    \item \textsc{UpdateValue}$(k, v)$ takes $O(\log n)$ work and depth.
    \item \textsc{UpdatePriority}$(k, p)$ takes $O(\log n)$ work and depth. 
    \item \textsc{Query}$(k)$ takes $O(\log n)$ work and depth.
    \item \textsc{Find}$(p)$ takes $O(\log n)$ work and depth.
    \item \textsc{NextWith}$(k, f)$ takes $O((q - k + 1) \log n)$ work with $O(\log^2 n)$ depth, where $q$ is the returned value of  \textsc{NextWith}$(k, f)$.
    \end{itemize}
\end{restatable}

Apart from the \textsc{NextWith} function, this data structure is essentially an array where elements are sorted in decreasing order of priority. \shortOnly{The proof of \Cref{lem:app_adjlist} is deferred to the full version.}

\fullOnly{\begin{proof}[Proof of \cref{lem:app_adjlist}]
We use the \textit{segment tree} data structure as defined by Lacki and Sankowski \cite{10.1145/2422436.2422468}, with the following distinction: Unlike the standard segment tree where all leaves and internal nodes are created in initialization, we create each node lazily only when we need access to a node in the subtree. Hence, a node whose corresponding interval was never accessed in any of the $\textsc{Initialize}$ or $\textsc{Query}$ functions will not be explicitly initialized. We store each element in the segment tree, indexed by its priority. Hence, an inorder traversal of the segment tree will list the elements in the increasing order of priority. Each segment tree node contains the count of values whose priority belongs to a corresponding interval. 

$\textsc{Initialize}$ function can be done by first sorting all the elements and using the standard segment tree operation of locating the position of index $p_i$, where it is done in parallel (by descending one depth at a time). $\textsc{UpdateValue}, \textsc{Query}, \textsc{Find}$ function is trivial. $\textsc{UpdatePriority}$ function is implemented by finding the $k$-th element by $\textsc{Query}(k)$, removing it (by decreasing the count of values by $1$), and adding a value in the specified position instead.

In the case of \textsc{NextWith}$(p, f)$, we devise an algorithm similar to exponential search. The search procedure will be divided into phases labeled with $0, 1, \ldots$. In phase $i$, we search for the first element among $\textsc{Query}(p), \textsc{Query}(p + 1), \ldots, \textsc{Query}(k, p + 2^i - 1)$ such that $f(\textsc{Query}(j)) = \mathsf{TRUE}$. This can be done in $O(2^i \log n)$ total work and $O(\log n)$ depth. If we find the first $j$ with $f(\textsc{Query}(j)) = \mathsf{TRUE}$, we return it. Otherwise, we add $2^i$ to $p$ and continue the search. If $p > k$, return $k + 1$.

Let $q$ be the value returned by the $\textsc{NextWith}$ function. The algorithm will terminate in at most $\lceil \log (q - p + 1) \rceil$ phases. Since $\lceil \log (q - p + 1) \rceil \le \lceil \log k \rceil = O(\log n)$, the search procedure has $O(\log^2 n)$ depth. The total work spent by the algorithm is $(2^0 + 2^1 + \ldots + 2^{\lceil \log (q - p + 1) \rceil}) \log n = O((q - p + 1) \log n)$.
\end{proof}}

\subsection{Maintaining Shallow Shortest Path Trees}
Another important ingredient we make use of is a batch-dynamic parallel algorithm for maintaining a BFS up to a certain small distance, in a graph with batches of edge deletions:
\estree*
The proof of \cref{thm:estree} is an adaptation of the classic sequential BFS of Even and Shiloach~\cite{shiloach1981line} to parallel batch-dynamic setting. We need to make several minor adjustments so that the algorithm can be implemented in low depth and could be modified to yield a $(2k-1)$-spanner in the forthcoming sections. 


In the initialization stage, we need to run a Breadth-First Search (BFS) from $s$ to initialize a shortest path tree of depth $L$. This can be done in a work-efficient way, as stated below:

\begin{restatable}{lemma}{AppBFS}
\label{lem:app_bfs}
    Given a unweighted directed graph $G = (V, E)$ where $|V| = n, |E| = m$, a source vertex $s \in V$, and a parameter $L \leq n$, there exists a deterministic algorithm that computes a size-$n$ array $\textsc{Dist}(*)$, where for each vertex $v$:
    \begin{itemize}
        \item If the shortest path from $s$ to $v$ exists and have a length $d \le L$, $\textsc{Dist}(v) = d$,
        \item otherwise, $\textsc{Dist}(v) = L+1$.
    \end{itemize}
    The algorithm takes $O(m \log n)$ work and $O(L \log n)$ depth.
\end{restatable}
\begin{proof}
    Let $S(i)$ be the set of vertices where the shortest path from $s$ is exactly of length $i$. Such a set is maintained with a binary search tree. Initially, we have $S(0) = \{s\}$. For each $i = 0, 1, \ldots, L - 1$, we can compute the set $S(i + 1)$ from the $S(i)$ in the following way: We iterate through all adjacent vertices that are not visited and add them into $S(i + 1)$. This can be done in $O(\log n)$ depth and $O(\log n)$ work per each out-edge of $S(i)$. Since each edge is considered in at most two phases of the above procedure, this yields $O(m \log n)$ total work. 
\end{proof}

We are ready to describe the initialization phase. We compute an array $\textsc{Dist}$ with \cref{lem:app_bfs}. Also, for each vertex $v$, we initialize two lists $\textsc{In}(v)$ and $\textsc{Out}(v)$, containing the in-edges that end at $v$ and out-edges that start at $v$, respectively. The list $\textsc{In}(v)$ is initialized using \cref{lem:app_adjlist}. Currently, the priorities are irrelevant and can be set arbitrarily - for example, the vertex index. The list $\textsc{Out}(v)$ is initialized as an array. 

For each vertex $v$ where $1 \le \textsc{Dist}(v) \le L$, we maintain a pointer $\textsc{Scan}(v)$ on a list $\textsc{In}(v)$. This pointer will satisfy the following invariant:

\paragraph{Invariant A1} $\textsc{Scan}(v)$ points to the first vertex of $\textsc{In}(v)$ with distance $\textsc{Dist}(v) - 1$ from source.

Let $T$ be the collection of edges pointed by $\textsc{Scan}(v)$. We can see that $T$ forms a shortest path tree rooted at $s$, consisting of vertices with distance at most $L$. Initially, $T$ can be computed by invoking a $\textsc{NextWith}$ function over $\textsc{In}(v)$ for each vertex $v$ with $1 \le \textsc{Dist}(v) \le L$, where we find the first incoming edge $(w \rightarrow v)$ in $\textsc{In}(v)$ where $\textsc{Dist}(w) = \textsc{Dist}(v) - 1$.

Consider the batch deletion query of edges $\{e_1, e_2, \ldots, e_k\}$, and let $e_i$ be the edge from vertex $u_i$ to vertex $v_i$. If the edge $e_i$ does not belong to $T$, we remove it from the data structure $\textsc{In}(v)$ and $\textsc{Out}(v)$ by marking it as an invalid edge. This can be done with a single call of the $\textsc{Set}$ operation per each edge, requiring an $O(1)$ work and depth. Hence, we are left only with edges from $T$.

Removing edges in $T$ is trickier since the value of $\textsc{Dist}$ and $\textsc{Scan}$ should be maintained correctly. For this, we proceed in \textit{phases} indexed as $i = 0, 1, \ldots, L$. For each phase, we maintain a set $U$ consisting of vertices $v$ with potentially incorrect $\textsc{Dist}(v)$ and $\textsc{Scan}(v)$. Throughout the algorithm, we maintain the following additional invariants:

\paragraph{Invariant A2} Before phase $i$, for all vertex $v \in V - U$ with distance at most $i$ has a correct distance.

\paragraph{Invariant A3} Before phase $i$, for all vertex $v \in U$, a vertex $w \in \textsc{In}(v)$ that precedes $\textsc{Scan}(v)$ do not have a distance less than $i$.

\paragraph{Invariant A4} Before phase $i$, for all vertex $v \in U$, their distance is at least $i$.

Maintaining these invariants for all phases is sufficient to maintain all distances correctly.

Initially (before phase $0$), we begin with $U = \{\}$. Then, we remove all edges from $T$ and update the pointer $\textsc{Scan}(v)$ to point to the next edge from $\textsc{In}(v)$, using the $\textsc{NextWith}$ function. This might cause the $\textsc{Scan}(v)$ to point to the end of the list, which does not correspond to any actual edges.

In phase $i$, all vertices in $v \in U$ will \textit{rescan} the list $In(v)$ from $\textsc{Scan}(v)$, and find if there exists a edge $(w \rightarrow v)$ with $\textsc{Dist}(w) = \textsc{Dist}(v) - 1$. By invariant A2, all scanned values of $\textsc{Dist}(w)$ are correct. If such an edge exists, we set the $\textsc{Scan}(v)$ pointer to such a vertex, updating the tree $T$. As a result, the values $\textsc{Dist}(v)$ and $\textsc{Scan}(v)$ are both correct, and no further work is required.

Otherwise, the distance from $s$ to $v$ is strictly greater than the current $\textsc{Dist}(v)$ value. This not only makes $\textsc{Dist}(v)$ incorrect, but all descendants of $v$ in the shortest path tree $T$ may potentially have an incorrect value of $\textsc{Dist}(v)$ and $\textsc{Scan}(v)$. Let $U_{new}$ be the set that will be used as $U$ in phase $i + 1$. We add $v$, and all its direct descendants in $T$ to $U_{new}$. These descendants can be found by simply iterating through $\textsc{Out}(v)$ and checking if the parent of a vertex is $v$. To reinitiate the search of the parent, we also set $\textsc{Scan}(v)$ to the head of the list $\textsc{In}(v)$. 

Finally, if $i + 1 \le L$, we put all vertices $v$ to $U_{new}$ such that $\textsc{Dist}(v) = i$ and their parent edge is removed. Before the next phase, we set $U = U_{new}$, and set the distance of all vertices in $U$ to $i + 1$. Below, we state a pseudo-code of our algorithm.

\begin{algorithm}\caption{Handling the deletion updates}\label{alg:central_delete}
\begin{algorithmic}[1]
\small
\State Remove all non-tree edges in $G$
\State Remove all tree edges in $G$ and set $\textsc{Scan}(v)$ to point the next edge in the list $\textsc{In}(v)$
\State $U = \{\}$ 
\For{$i\gets 0, 1, \ldots, L$}
\State  $U_{new} = \{\}$
\Comment{Starting the phase $i$.}
\For{$v \in U$ \text{in parallel}}
\State $\textsc{Scan}(v) = \textsc{In}(v).\textsc{NextWith}(\textsc{Scan}(v), f(w) = (\textsc{Dist}(w) = \textsc{Dist}(v) - 1))$
\If{\textsc{Scan}(v) = \textsc{In}(v).size + 1}
\State $\textsc{Scan}(v)=1$
\State $U_{new}.\textsc{Add}(\{v\})$
\State $U_{new}.\textsc{Add}(\textsc{Child}_T(v))$ \Comment{$\textsc{Child}_T(v)$ is a list of child of $v$ in the BFS tree $T$.}
\EndIf
\EndFor
\State Add all vertex $v$ to $U_{new}$ such that $\textsc{Dist}(v) = i$ and the parent edge is removed
\State $U := U_{new}$
\For{$v \in U$ \text{in parallel}}
\State $\textsc{Dist}(v) = i + 1$
\EndFor
\EndFor
\end{algorithmic}
\end{algorithm}

We analyze the depth and work requirements of \cref{alg:central_delete}. For the initialization, the algorithm requires a single call of \cref{lem:app_bfs}, and a total $n$ call of $\textsc{NextWith}$ function, totaling to $O(L \log n + \log^2 n)$ depth and $O(m \log n)$ work. For the edge deletion updates, the depth is bounded by $O(L \log^2 n)$, as each phase can be implemented in $O(\log^2 n)$ time with our data structures. The algorithm's work is dominated by the set operation incurred by $U$ and the utilization of the $\textsc{NextWith}$ function to move the pointer $\textsc{Scan}(v)$ to new parents. The total movement made by all of the pointers of $\textsc{Scan}(v)$ is at most $O(Lm)$. As a result, the total work required by $\textsc{NextWith}$ function is $O(Lm \log n)$. An insertion in the set $U$ only occurs when the pointer $\textsc{Scan}(v)$ has moved - hence this is also bounded by $O(Lm)$. Each insertion requires $O(\log n)$ work; hence, the algorithm requires $O(Lm \log n)$ work in total, which is $O(L \log n)$ amortized work per deletion.

\subsection{Decremental $(2k-1)$-Spanners}\label{sec:decremental-mpvx15}
In this section, we prove the following lemma, which is essentially \cref{thm:mainnear} restricted to decremental updates.

\begin{lemma}\label{lem:mainnear}
    There is a randomized parallel batch-dynamic decremental data structure which, given an unweighted graph $G = (V, E)$ where $|V| = n, |E| = m$, maintains a $(2k-1)$-spanner of $O(n^{1 + 1/k})$ expected number of edges. Specifically, the algorithm supports the following interfaces:

    \begin{itemize}
        \item After the initialization, the algorithm returns a set of edges forming a $(2k-1)$-spanner of the given graph, which has $O(n^{1+1/k})$ expected number of edges.
        \item After each edge insertion and deletion update, the algorithm returns a pair of edge sets $(\delta H_{ins}, \delta H_{del})$, representing the set of edges that are newly inserted or deleted into the $(2k-1)$-spanner. The amortized size of $|\delta H_{ins}| + |\delta H_{del}|$ is at most $O(k \log n)$ per edge in expectation.
    \end{itemize}

    The algorithm takes:
\begin{itemize}
    \item for initialization, $O(m \log n)$ expected work, and $O(\log^2 n)$ worst-case depth,
    \item for any batch of edge deletions, $O(k \log^2 n)$ expected amortized work per deleted edge, and $O(k \log^2 n)$ worst-case depth for the entire batch.
\end{itemize}
The above statements hold with high probability against an oblivious adversary.
\end{lemma}

Our decremental algorithm is based on the \textit{exponential start time clustering} introduced by \cite{miller2015improved}. In the algorithm of \cite{miller2015improved}, each vertex picks a real number $\delta_u$ from the exponential distribution with parameter $\frac{\log (10n)}{k}$. Recall the exponential distribution with parameter $\beta$, denoted $\text{Exp}(\beta)$, has a density $\beta \cdot e^{-\beta x}$ for $x \geq 0$. In guaranteeing a $(2k-1)$ stretch, their algorithm is a Monte Carlo algorithm with constant failure probability. In \cref{alg:mpvx15}, we present a \textit{Las Vegas modification} - in lines 1-3, if we have a vertex $u$ with $\delta_u \geq k$, we regenerate all values $\delta_u$, unlike the version in \cite{miller2015improved} where they generate the values once. Hence, our algorithm returns the spanner with $(2k-1)$-stretch with high probability. 

\begin{algorithm}[H]
\caption{Algorithm 2 of \cite{miller2015improved}, modified to guarantee $(2k-1)$ stretch}\label{alg:mpvx15}
\begin{algorithmic}[1]
\small
\State $\delta_u = k$ for all $u \in V$.
\While{$\max \delta_u \geq k$}
\State For each vertex $u$, pick $\delta_u$ independently from the exponential distribution Exp$(\frac{\log (10 n)}{k})$. 
\EndWhile
\State Create clusters by assigning each $v \in V$ to $u = \arg \min_{u \in V}\{dist(u, v) - \delta_u\}$, if $v = u$ we call it a center of its cluster. Let $\textsc{Cluster}(v)$ be the cluster to which $v$ belongs. 
\State Construct a forest $H$, where each component is a spanning tree on each cluster rooted at its center.
\For{$v \in V$}
\State $C = \{\textsc{Cluster}(w) \mid \textsc{Cluster}(w) \neq \textsc{Cluster}(v), (v, w) \in E\}$
\For{$c \in C$}
\State Pick an edge $(v, w) \in E$ where $\textsc{Cluster}(w) = c$.
\State Add $(v, w)$ to $H$.
\EndFor
\EndFor
\State \Return $H$.
\end{algorithmic}
\end{algorithm}

To prove that \cref{alg:mpvx15} returns a correct spanner, we cite the analysis of \cite{miller2015improved}, due to Elkin and Neiman  \cite{elkin2018efficient}.

\begin{lemma}[Lemma 3, 6 of \cite{elkin2018efficient}]\label{lem:elkinneiman2}
With probability $0.9$, line 3 of \cref{alg:mpvx15} will generate $\delta_u$ such that $\max_{u\in V} \delta_u < k$. Consequently, $H$ is a spanner of stretch $(2k-1)$.
\end{lemma}

\begin{lemma}[Lemma 2 of \cite{elkin2018efficient}]\label{lem:elkinneiman1}
$H$ has $O(n^{1+1/k})$ edges in expectation.
\end{lemma}
\begin{proof}
    The expectation in Lemma 2 of \cite{elkin2018efficient} is taken without lines 1-3 of our algorithm, which samples under the condition $\max \delta_u < k$. However, if this conditioning increases the expected number of edges by more than a $\frac{10}{9}$ factor, we obtain a contradiction to the original statement. 
\end{proof}

Hence, it suffices to adapt \cref{alg:mpvx15} under decremental updates. Let $\delta_u = d_u + f_u$, where $d_u \geq 0$ is an integer and $f_u$ is a real number in range $[0, 1)$. We define two objects: the first one, based on the integer $d_u$, and the other, based on the real number $[0, 1)$.

The first object is an auxiliary directed graph $G^\prime = (V^\prime, E^\prime)$. Here, $V^\prime = V \cup \{p_0, p_1, \ldots, p_{t-1}\}$ for $t = (\max_{v} d_v) + 1$, and $E^\prime$ is constructed in a following way:
\begin{itemize}
    \item For each edge $(u, v) \in E$, add two directed edges $(u \rightarrow v), (v \rightarrow u)$ in $E^\prime$.
    \item For each $0 \le i \le t - 2$, add a directed edge from $p_i$ to $p_{i + 1}$ in $E^\prime$.
    \item For each vertex $v$ from $G$, add a directed edge from $p_{t-1-d_v}$ to $v$ in $E^\prime$.
\end{itemize}

The second object is a \textit{priority} $\textsc{Priority}(v)$, which is a permutation of $V$ sorted in increasing order of $f_u$. This permutation can be obtained by sorting, which takes $O(n \log n)$ work and $O(\log n)$ depth. For analysis, we assume every $f_u$ is distinct, which holds with high probability. 

For each vertex $v \in V$, any vertex $u$ that minimizes $dist(u, v) - \delta_u$ also minimizes $dist(u, v) - d_u$. Consider the shortest path tree of $G^\prime$ rooted in vertex $p_0$. The shortest distance from $p_0$ to a vertex $v \in V$ equals to $t - d_u + dist(u, v)$, where $u$ is the lowest vertex (closest to the root $p_0$) in the shortest path not part of a path $\{p_0, \ldots, p_{t-1}\}$. Hence, this $u$ minimizes the value $dist(u, v) - d_u$. From all possible $u$ that could be in the shortest path from $p_0$ to $v$, we choose the one that maximizes $f_u$ - in other words, maximizes the priority $\textsc{Priority}(u)$. We can see that this is the value $\textsc{Cluster}(v)$ computed in line 4 of \cref{alg:mpvx15}. 

We use the parallel Even-Shiloach of \cref{thm:estree} in $G^\prime$ with $s = p_0, L = t$ to maintain these invariants. This gives a shortest-path tree of depth $t$ that contains every vertex of $G^\prime$. The cluster of each vertex is declared recursively, as following: For a vertex $v$ whose parent $par(v)$ in the shortest path tree is in $\{p_0, \ldots, p_{t-1}\}$, we declare $\textsc{Cluster}(v) = v$. Otherwise, we follow the cluster of the parent, as $\textsc{Cluster}(v) = \textsc{Cluster}(par(v))$. 

Each deletion update may change the parent of a vertex in the shortest path tree, which can be traced by seeing if $\textsc{Scan}(v)$ is updated. Hence, after the deletion updates, we check such a set of vertices and update the cluster information. This may cause the descendants in the shortest path tree to change their clusters, so we recursively change them.

Unfortunately, the above scheme only guarantees the cluster $\textsc{Cluster}(v)$ to minimize $dist(u, v) - d_u$, but not $dist(u, v) - \delta_u$ - we need to take account of the priority. For this, while initializing the Even-Shiloach data structure of \cref{thm:estree}, we have to carefully initialize the data structure of \cref{lem:app_adjlist} on $\textsc{In}(v)$. When initializing the data structure $\textsc{In}(v)$ for each $V$, we supply a priority value for an edge $(w \rightarrow v)$ as:
\begin{itemize}
    \item $\textsc{Priority}(\textsc{Cluster}(w))$ if $w \in V$,
    \item $\textsc{Priority}(v)$ otherwise (if $w \in \{p_0, \ldots, p_{t-1}\}$).
\end{itemize}

This keeps every edge in $\textsc{In}(v)$ sorted in decreasing order of priority. As a result, $\textsc{Scan}(v)$ will point to the vertex of $\textsc{In}(v)$ such that its distance from source is $\textsc{Dist}(v) - 1$, and maximizes the value $\textsc{Priority}(\textsc{Cluster}(w))$. By induction on the length of the shortest path, together with the invariant A1 of \cref{thm:estree}, we can see that this rule considers all possible candidates of $\textsc{Cluster}(v)$ and takes the one that maximizes the priority.

With all the invariants in place, we discuss the implementation of our algorithm. We assume the graph contains no duplicate edges (which we can guarantee using hash tables). In the initialization phase, we sample $\delta_u$ from the exponential distribution (retrying while $\max \delta_u < k$), generate the priority permutation, and then initialize the Even-Shiloach of \cref{thm:estree} where $\textsc{In}(v)$ is equipped with a priority in a way we discussed earlier. Additionally, we maintain a hash table $\textsc{InterCluster}$ that maps a pair of integers $(v, c)$ to another hash table. That is, $\textsc{InterCluster}$ is a hash table of hash tables. Each entry of $\textsc{InterCluster}$, which is a hash table $\textsc{InterCluster}[(v, c)]$, maintains the set of edges whose one endpoint is a vertex $v$ and the other is in cluster $c$. 

The spanner $H$ consists of the intra-cluster part, which comes from the edges of the shortest path tree of $G^\prime$, and the inter-cluster part, which comes from the hash table. To obtain the inter-cluster edges, we pick a single edge from each $\textsc{InterCluster}[(v, c)]$ that is not empty and satisfies $c \neq \textsc{Cluster}(v)$. 

Given a set of edge deletions, we need to apply them to the parallel Even-Shiloach data structure, which can be easily done by invoking the deletion update of \cref{thm:estree}. Tracking the set of updated intra-cluster edges is also straightforward. Maintaining the valid $\textsc{Cluster}(v)$ values and associated priority is the real challenge. To maintain the valid priority values for $\textsc{In}(v)$ among the changes incurred to the value $\textsc{Cluster}(v)$, we implement a recursive procedure similar to \cref{alg:central_delete}. Suppose the vertex $v$ acquires the new $\textsc{Cluster}(v)$ value. Then, we iterate through all vertices $w \in \textsc{Out}(v)$ in parallel, and update the priority value stored in $\textsc{In}(w)$ corresponding to the edge $(v \rightarrow w)$ using the function $\textsc{Find}$ and $\textsc{UpdatePriority}$. Then, for each of these vertices, we see if they need to acquire new cluster values: Since the value $\textsc{Priority}(\textsc{Cluster}(v))$ is nondecreasing unless the distance changes, we can detect this with a single $\textsc{NextWith}$ call. If they need to acquire new cluster values, we change them and proceed recursively. Whenever the value $\textsc{Cluster}(v)$ changes, we can iterate, in parallel, through all edges incident on $v$ and update the entries corresponding to that edge in the hash table $\textsc{InterCluster}$ correspondingly. 

This ends the description of our algorithm. In \cref{lem:elkinneiman1} and \cref{lem:elkinneiman2}, we showed that the algorithm correctly returns the desired spanner, so it suffices to argue that our algorithm is efficient. We present a key lemma for the analysis of our algorithm:

\begin{lemma}\label{lem:cluster2}
    For each vertex $v$, the expected number of times $\textsc{Cluster}(v)$ changes is at most $2 t \log n$ if the invariant above holds. 
\end{lemma}

\shortOnly{The proof of \Cref{lem:cluster2} is deferred to the full version.}
\fullOnly{
\begin{proof}
Assuming that every $f_u$ is distinct, we prove that $\textsc{Priority}$ is a uniformly random permutation, independent of the choice $d_u$. This comes from the memoryless property of the exponential distribution. For any vertex $u$ with fixed $d$, the probability of $f_u \geq f$ is:

\begin{align*}
Pr((\delta_u \geq f + d) | ((\delta_u \geq d) \land \lnot (\delta_u \geq d + 1))\\
= \frac{ Pr((\delta_u \geq f + d) \land ((\delta_u \geq d) \land \lnot (\delta_u \geq d + 1))}{Pr((\delta_u \geq d) \land \lnot (\delta_u \geq d + 1))}\\
= \frac{e^{-\beta (f + d)} - e^{-\beta (d + 1)}}{e^{-\beta d} - e^{-\beta (d+1)}} = \frac{e^{-\beta f} - e^{-\beta}}{1 - e^{-\beta}}
\end{align*}
where $\beta = \frac{\log (10 n)}{k}$ is fixed. Hence, we can assume every $f_u$ comes from an identical independent distribution, which proves that the ordering between them is uniformly random.

For each vertex $v$, assume that the shortest path length from the source vertex $p_0$ is currently $d$. Consider the set of vertices $U = \{u_1, u_2, \ldots, u_k\}$ where $d = t - \delta_{u_i} + dist(u_i, v)$. In other words, those are the set of vertices that could be the valid cluster of $v$. Additionally, we assume that $U$ is ordered in the following way: As the edges in the graph are removed, each $u_i$ will cease to belong in $U$ and not be a valid candidate for the cluster. We order them so that the time it ceases to satisfy the condition is nonincreasing. For example, $u_1$ will maintain the condition $d = t - \delta_{u_i} + dist(u_i, v)$ for the longest time.

For each $u_i$, the probability of the $u_i$ being ever chosen as the cluster of $v$, in other words, $u_i = \textsc{Cluster}(v)$, is $\frac{1}{i}$. To see this, note that $u_i$ will ever be chosen only if for all $1 \le j \le i$, $\textsc{Priority}(u_j) \le \textsc{Priority}(u_i)$: Otherwise, the one with higher priority will be selected as a cluster, and will remain so until $u_i$ cease to be a valid cluster. As we assumed the $\textsc{Priority}$ to be a random permutation, such an event will happen with probability $\frac{1}{i}$. By linearity of expectation, it follows that the expected number of $u_i$ being ever chosen as the cluster of $v$ is at most $\sum_{i = 1}^{k} \frac{1}{i} \le 2 \log k \le 2 \log d$. 

The shortest path length from the source vertex $p_0$ will only increase and is bounded by $t$. As a result, the expected total number of times $\textsc{Cluster}(v)$ changes is at most $2t \log n$.
\end{proof}}

\begin{proof}[Proof of \cref{lem:mainnear}]
We analyze the depth and work requirements of our algorithm. Obtaining $\delta_u$ and its associated priority takes $O(\log^2 n)$ depth and $O(n \log n)$ work with high probability. The parallel Even-Shiloach algorithm of \cref{thm:estree} takes $O(\log^2 n)$ depth and $O(m \log n)$ work to initialize and report the set of edges forming the initial clusters. Then, each edge deletion takes $O(k \log^2 n)$ depth and $O(k \log n)$ amortized work. Our additional procedure of computing the $\textsc{Priority}$ and assigning it in the initialization stage takes $O(\log n)$ work and depth per edge, which is dominated by the work/depth of \cref{thm:estree}. A chain of cluster updates can be processed in the same $O(k \log^2 n)$ depth, and each update requires $O(d \log n)$ work, where $d$ is the degree of the updated vertex. By \cref{lem:cluster2}, this totals to $O(m k \log^2 n)$ work, which is $O(k \log^2 n)$ amortized work.

Next, we analyze the size of the edges $|\delta H_{ins}| + |\delta H_{del}|$. In the parallel Even-Shiloach tree, each edge ceases to form a shortest path tree only when it is deleted from the graph. Hence, a total of $O(m)$ intra-cluster edges are updated in the algorithm. Inter-cluster edges may enter or leave the spanner only if one of the endpoints has changed the cluster. By \cref{lem:cluster2}, this happens at most $O(k\log n)$ times, giving a bound of $O(m k \log n)$, which implies $O(k \log n)$ amortized expected size per edge update.
\end{proof}

\subsection{Fully-Dynamic $(2k-1)$-Spanners}\label{sec:full-mpvx15}
We provide a reduction from the fully-dynamic algorithm for $(2k-1)$-spanners to the decremental algorithm, thus allowing us to use our decremental algorithm and ultimately prove our main theorem of this section, i.e., \Cref{thm:mainnear}. Our reduction is very similar to that of Baswana et al. \cite{baswana2008fully}, which begins with the following observation:

\begin{observation}[Observation 5.2 of \cite{baswana2008fully}]
    Given an undirected graph $G = (V, E)$, let $E_1, \ldots, E_j$ be the partition of the set of edges $E$, and let $H_1, \ldots, H_j$ be respectively the $t$-spanners of subgraphs $G_1 = (V, E_1), \ldots, G_j = (V, E_j)$. Then, $\cup_i H_i$ is a $t$-spanner of the original graph $G = (V, E)$.
\end{observation}

Any problem that satisfies such a property is called \textit{decomposable}, and a classical result of \cite{BENTLEY1980301} states that any decremental algorithm on a decomposable problem can be transformed to a fully dynamic algorithm of similar amortized complexity. In the proof of \cref{thm:mainnear}, we recite the result of \cite{BENTLEY1980301} and verify that their reduction also works in the batch-parallel setting.

We are ready to provide the proof of \Cref{thm:mainnear}.

\begin{proof}[Proof of \cref{thm:mainnear}]
We assume the graph does not contain duplicate edges throughout the operations. These can be assumed by filtering out duplicate edges using the hash tables.

The fully dynamic algorithm maintains the partition of edges $E = E_0 \cup E_1 \cup \ldots \cup E_b$, where $b \leq O(\log n)$. Additionally, the algorithm maintains a global hash table $\textsc{Index}(e)$ which, given an edge $e \in E$, maintains the index $i$ where $e \in E_i$. Let $l_0$ be the smallest integer such that $2^{l_0} \geq n^{1 + 1/k}$. Each partition $E_i$ satisfies the following invariant:

\paragraph{Invariant B1} $|E_i| \leq 2^{i + l_0}$.

Under the invariant B1, it is trivial to maintain $E_0$, as we can put all edges in the spanner. For any other $E_i$ with $i \geq 1$, we use the decremental $(2k-1)$-spanner algorithm of \cref{lem:mainnear}.

Initially, the algorithm locates the smallest $j$ such that $|E| \leq 2^{j + l_0}$ and initializes the decremental structure on $E_j$ with the edge set $E$. All other partitions are empty. 

Consider the edge deletion update. The algorithm uses the hash table $\textsc{Index}$ to find the partition each edge belongs to. With this information, the algorithm deletes the edge in its respective partition using the decremental algorithm. One can easily see that this procedure runs in the same work/depth bound as in \cref{lem:mainnear} and does not break any invariants.

Consider the edge insertion update, where we try to insert a set of edges $U$. We first divide $U$ into a partition of $b+2$ sets $U = U_r \cup U_0 \cup U_1 \cup \ldots \cup U_b$, such that for each $0 \le i \le b$. Set $U_i$ is either empty or $|U_i| = 2^{l_0 + i}$, and we have $0 \le |U_r| < 2^{l_0}$. Note that the size of $U$ uniquely determines which $U_i$ are empty.

We will then process the sets, in the order of $U_b, U_{b-1}, \ldots, U_0$, in the following manner: Assuming that $U_i$ is not empty (otherwise we ignore), let $j \geq i$ be the smallest $j$ such that $E_j$ is empty. We initialize $E_j = (U_i \cup E_i \cup E_{i + 1} \cup \ldots \cup E_{j-1})$, and empty out all the set $E_i, E_{i+1}, \ldots, E_{j-1}$. Finally, for the set $U_r$, we check if $|U_r| + |E_0| \leq 2^{l_0 + i}$. If the condition holds, we simply add all the edges $U_r$ to $E_0$ and the corresponding spanner. Otherwise, we find the first $j \geq 0$ such that $E_j$ is empty, initialize $E_j = (U_r \cup E_0 \cup E_1 \cup \ldots \cup E_{j-1})$, and empty out all the set $E_0, E_1, \ldots, E_{j-1}$. 

We need to prove three things for the edge insertion updates: First, check if the invariants are broken in the procedure. Second, ensure we always find such $j \geq i$ where $E_j$ is empty. Finally, prove that the given procedure has a polylogarithmic depth and span. The first point can be easily seen, since for nonempty $U_i$, we have $|U_i| + \sum_{k = i}^{j-1} |E_k| \leq 2^{l_0 + i} + \sum_{k = i}^{j-1} 2^{l_0 + k} = 2^{l_0 + j}$. For the set $U_r$, the same argument holds. It remains to prove the second and third points.

For the second point, let $\Phi = \sum_{i = 0}^{b} (\mathbf{1}[E_i \neq \emptyset] 2^i)$. In other words, $\Phi$ is the sum of $2^i$ for each nonempty $E_i$. Initially, $\Phi \leq n^{1 - 1/k}$ as there are at most $n^2$ edges. The deletion update will never increase the $\Phi$. Suppose that we are doing the insertion update. When processing the nonempty $U_i$, we increase the $\Phi$ by $2^j$, while decreasing it by $2^i + 2^{i+1} + \ldots + 2^{j-1}$. Hence, each $U_i$ will increase the $\Phi$ by exactly $2^i$. Similarly, $U_r$ will increase the $\Phi$ by at most $1$. Hence, an insertion update of size $U$ will increase the $\Phi$ by at most $\lceil \frac{|U|}{2^{l_0}} \rceil$. Hence, as long as there are at most a polynomial number of updates, we can assume that $\Phi \leq poly(n)$, leaving an empty $E_i$ for some $i = O(\log n)$. If there are more than a polynomial number of updates, we can restart the algorithm once every $O(n^3)$ updates, which can be done in low depth and requires $O(1)$ amortized work.

As a final point, we review the work, depth, and the recourse bounds (i.e., the amortized size of $|\delta H_{ins}| + |\delta H_{del}|$). Note that whenever the edge is a part of the initialization, either it is a fresh new edge or the value $\textsc{Index}(e)$ has increased. As a result, each edge belongs to at most $O(\log n)$ decremental instances. 

The insertion takes $O(k \log^2 n)$ depth and $O(\log^2 n)$ work, as a result: The work has an extra $O(\log n)$ factor, but the depth remains as is, since we do not need to initialize each $E_j$ sequentially - rather, we compute the set of edges to initialize after processing $U_i$, and then initialize each of them in parallel. For the deletion and initialization cases, the bound is identical with \cref{lem:mainnear}. Finally, by \cref{lem:mainnear}, the total recourse of each decremental instance is $O(m k \log n)$, so the total recourse of this fully-dynamic instance is $O(m k \log^2 n)$.  
\end{proof}

\section{Dynamic Contractions, and Sparse Spanners}
Here, we present our batch-dynamic sparse spanner with a linear number of edges, achieving this via iterative contractions, thus proving \cref{thm:maintrue}.

\subsection{The Contraction Procedure}\label{sec:contraction}
We describe the procedure $\textsc{Contract}(G, x)$, which takes a simple graph $G = (V, E)$ with $n$ vertices and a real parameter $x$, and returns a subproblem of size $\frac{|V|}{x}$. Formally, we state the following:

\begin{lemma}\label{lem:contraction}
    There is a procedure $\textsc{Contract}(G, x)$ which takes a simple graph $G = (V, E)$ with $n$ vertices and a parameter $2 \le x \le O(\log n)$, and returns a tuple $(G^\prime, H, f)$ such that the following holds with high probability:
\begin{itemize}
    \item $G^\prime = (V^\prime, E^\prime)$ is a graph, $f : V \rightarrow V^\prime \cup \{\bot\}$ is a function, $H \subseteq E$ is a subset of edges.
    \item $V^\prime$ is not empty.
    \item For any $y \in V^\prime$, $f(y) = y$ holds.
    \item $E[|V^\prime|] = \frac{n}{x}$.
    \item $E[|H|] = O(nx)$.
    \item For any $L$-spanner of $G^\prime$ denoted as $H^\prime$, we can find a $(3L+2)$-spanner of size $|H^\prime| + |H|$, that contains all the edges of $H$.
\end{itemize}
\end{lemma}

Towards giving an intuition for the proof of \Cref{lem:contraction}, we first describe the simple sequential version of $\textsc{Contract}(G, x)$. We will later describe ingredients to adapt the $\textsc{Contract}(G, x)$ into a parallel batch-dynamic model that will be used to prove \cref{thm:maintrue}.

We begin by computing the set of vertices $D \subseteq V$, uniformly sampled from $V$ with probability $\frac{1}{x}$. For all the vertices $v \in V - D$, we compute the \textit{head} of each vertex $\textsc{Head}(v)$ in the following way:

\begin{itemize}
\item If $v$ has a neighbor $w$ that is sampled, take any such $w$ and declare $\textsc{Head}(v) = w$.
\item Otherwise, $\textsc{Head}(v) = \bot$
\end{itemize}

Using the information $\textsc{Head}(v)$, the graph $G^\prime = (V^\prime, E^\prime)$ is constructed by removing all vertices with $\textsc{Head}(v) = \bot$, contracting each pair of vertices $(v, \textsc{Head}(v))$, and remove all self-loops or duplicate edges to make it a simple graph. Formally, we set $V^\prime$ as $D$, and for all pair of $(u^\prime, v^\prime)$ such that $u^\prime \neq v^\prime, \{u^\prime, v^\prime\} \in V^\prime$, we put an edge between $u^\prime$ and $v^\prime$ in $E^\prime$ if there exists an edge $(u, v)$ in $E$ such that $\textsc{Head}(u) = u^\prime, \textsc{Head}(v) = v^\prime$. 

The set $H$ consists of two types of edges. First, all edges $(u, v) \in E$ such that either $\textsc{Head}(u) = \bot$ or $\textsc{Head}(v) = \bot$ will be added into $H$. Next, for all vertices with $\textsc{Head}(v) \neq v, \textsc{Head}(v) \neq \bot$, an edge between $\textsc{Head}(v)$ and $v$ will be added, which definitely exists in $E$. Finally, we set $f = \textsc{Head}$. 

\begin{algorithm}\caption{Simple Sequential Implementation of $\textsc{Contract}(G, x)$}\label{alg:contract1}
\begin{algorithmic}[1]
\small
\State $D = \{\text{vertices of } V \text{ sampled with probability }\frac{1}{x}\}$
\State $H = \emptyset$
\For{$v \in V - D$}
\If{There exists a neighbor $w\in D$ of $v$}
\State{$\textsc{Head}(v) = w$}
\State{Add edge $(v, w)$ into $H$}
\EndIf
\If{No such neighbor $w$ of $v$ exists}
\State{$\textsc{Head}(v) = \bot$}
\State{Add all edges incident to $v$ into $H$}
\EndIf
\EndFor
\State {$V^\prime = D$}
\State {$E^\prime = \{(u^\prime, v^\prime) \mid u^\prime \neq v^\prime, u^\prime \in D, v^\prime \in D, \exists (u, v) \in E \text{ such that } \textsc{Head}(u) = u^\prime, \textsc{Head}(v) = v^\prime\}$}
\State \Return {$(G^\prime = (V^\prime, E^\prime), H, f = \textsc{Head})$}
\end{algorithmic}
\end{algorithm}

This finishes the description of the procedure $\textsc{Contract}(G, x)$. In \cref{alg:contract1}, we show a pseudo-code of the procedure described before.

\fullOnly{
\begin{proof}[Proof of \cref{lem:contraction}]
    By definition, $G^\prime, f, H$ are objects of a given type, and $f(y) = y$ holds for all $y \in V^\prime$. The expected number of vertices in $V^\prime$ is $\frac{n}{x}$. The probability of $V^\prime$ being empty is $\exp(-\Omega(\frac{n}{x}))$, which is negligible under our parameters. We prove that the expected number of edges in $H$ is at most $O(nx)$. Recall that $H$ consists of two types of edges: edges that are incident to a vertex with $\textsc{Head}(v) = \bot$, and edges that connect $v$ to $\textsc{Head}(v)$. Clearly, there are at most $n$ edges of the second type, and it suffices to bound the first type of edges. Since every edge of the first type is incident to at least one vertex with $\textsc{Head}(v) = \bot$, the number of such edges are upper bounded to $\sum_{\textsc{Head}(v) = \bot} deg(v)$, where $deg(v)$ is the degree of vertices. For a vertex of degree $d$, the probability that such vertex will have $\textsc{Head}(v) = \bot$ is $(1 - \frac{1}{x})^{d + 1}$, as $v$ and all its neighbor should not be sampled. As a result, the expected value of $\mathbf{1}[\textsc{Head}(v) = \bot] \cdot deg(v)$ is $(1 - \frac{1}{x})^{deg(v)+1} deg(v)$, which is upper bounded by $O(x)$ for any possible value of $deg(v)$. From the linearity of expectation, we can deduce that the expected value of $\sum_{\textsc{Head}(v) = \bot} deg(v)$ is $O(nx)$.
    
    Finally, we show an algorithm for finding a $(3L+2)$-spanner of size $|H^\prime| + |H|$ that contains all edges of $H$, given the $L$-spanner $H^\prime$. For edge $(u^\prime, v^\prime) \in E^\prime$, we pick any $(u, v) \in E$ such that $\textsc{Head}(u) = u^\prime, \textsc{Head}(v) = v^\prime$, and say $(u, v)$ is \textit{corresponding} to $(u^\prime, v^\prime)$. From line 11 of \cref{alg:contract1}, it is clear that every edge in $E^\prime$ can pick a corresponding edge in $E$. Our spanner consists of two types of edges: One, being all the edges of $H$, and another, being the set of corresponding edges of $H^\prime$. Clearly, we have a spanner of size $|H^\prime| + |H|$ that contains all the edges of $H$. It only remains to show that the stretch is at most $3L+2$. Two types of edges $(u, v) \in E$ are omitted in the spanner. One is the edges such that $\textsc{Head}(u) = \textsc{Head}(v)$, and another is the edges where $(\textsc{Head}(u), \textsc{Head}(v)) \in E^\prime$ but is not corresponding to $(\textsc{Head}(u), \textsc{Head}(v))$. For the first type of edges, one can pass through $u$ and $v$ via at most $2$ edges in $H$. For the other type of edges, as $H^\prime$ is a $L$-spanner of $G^\prime$, one can move through $\textsc{Head}(u)$ and $\textsc{Head}(v)$ in $G^\prime$ using at most $L$ edges in $H^\prime$. Replacing each of the edges in the path with the corresponding edges, we need to add $2(L+1)$ edges in $H$, which is used to move between vertices that share the same $\textsc{Head}(v)$. This amounts to $3L+2$ edges in total, proving the $3L+2$ upper bound on the stretch.
\end{proof}}

\subsection{Nested Contractions for Sparse Spanners}\label{sec:contraction_nest}
\cref{thm:mainnear} states that we can maintain an $O(\log n)$-spanner of at most $O(n \log n)$ edges dynamically. As we are not yet ready to discuss the full proof of \cref{thm:maintrue}, we assume that a black box can compute an $O(\log n)$-spanner of at most $O(n \log n)$ edges, and we want to compute a \strtch-spanner of at most $O(n)$ edge without modifying the black box, using some static sequential procedure that can be made parallel batch-dynamic. 

Our strategy is to apply the procedure $\textsc{Contract}(G, x)$ recursively, with appropriate parameters, to reduce the number of vertices by an $\log n$ factor, while losing only a little in the stretch. Specifically, suppose that there is a sequence $\{x_0, x_1, \ldots, x_{L-1}\}$ such that the following holds:

\begin{itemize}
    \item $x_i \geq 2$
    \item $\prod_{i = 0}^{L-1} x_i \geq \Omega(\log n)$
    \item $L \leq O(\log \log \log n)$
    \item $\sum \frac{x_i}{x_0 x_1 \ldots x_{i-1}} \leq O(1)$
\end{itemize}

Suppose we are given such a sequence. Then, the algorithm $\textsc{NestedConstract}(G, \{x_0, \ldots, x_{L-1}\})$ computes a spanner with stretch \strtch \text{ and} $O(n)$ edges. In particular, the algorithm proceeds in \textit{levels} indexed by $0, 1, \ldots, L-1, L$: Let $G_0$ be the input graph. In the $i$-th level where $i < L$, the algorithm calls $\textsc{Contract}(G_i, x_i)$, which will return a tuple that we assign as $(G_{i+1}, H_i, f_i)$. In the $L$-th level, the algorithm calls a black-box algorithm to return an $O(\log |V_L|)$-spanner of $O(|V_L| \log |V_L|)$ edges, where $G_L = (V_L, E_L)$. Denoting this spanner as $H_L$, we rewind the level from $L-1, L-2, \ldots, 0$, and compute the spanner $H_{L-1}, H_{L-2}, \ldots$. For completing $H_i$, we take the spanner of the $i+1$-th level $H_{i+1}$, and add each of the corresponding edges to $H_i$. \cref{alg:contract2} describes the implementation of the $\textsc{NestedContraction}$ in the $i$-th level:

\begin{algorithm}\caption{$\textsc{NestedContract}(G_i = (V_i, E_i), \{x_i, \ldots, x_{L-1}\})$}\label{alg:contract2}
\begin{algorithmic}[1]
\small
\If{$i = L$}
\State $H_L = $ a $O(\log |V_L|)$-spanner of $G$ with $O(|V_L| \log |V_L|)$ edges
\State \Return $H_L$
\EndIf
\State $(G_{i+1}, H_i, f_i) = \textsc{Contract}(G_i, x_i)$
\State $H_{i+1} = \textsc{NestedContract}(G_{i+1}, \{x_{i+1}, \ldots, x_L\})$
\State Add the corresponding edges of $H_{i+1}$ to $H_i$ using $f_i$ and $E_i$
\State \Return $H_i$
\end{algorithmic}
\end{algorithm}

By \cref{lem:contraction}, each level of the algorithm takes a $(t-1)$-spanner and returns a spanner with stretch $(3t-1)$. Hence, the algorithm $\textsc{NestedConstract}(G, \{x_0, \ldots, x_{L-1}\})$ returns a spanner of $G$ with stretch \strtch. Additionally, each level of the algorithm takes a graph that has $O(\frac{n}{x_0 \ldots x_{i-1}})$ vertices in expectation, and the number of edges in $H_i$ will be \\$O(\frac{n x_i}{x_0 \ldots x_{i-1}})$ in expectation. Since $\sum \frac{x_i}{x_0 x_1 \ldots x_{i-1}} \leq O(1)$, summing them for all $i$ will yield a spanner with $O(n)$ edges. 

By \cref{alg:contract2}, it suffices to construct a sequence $x_0, x_1, \ldots, x_{L-1}$ that satisfies the assumed condition. We show an explicit construction of a sequence in the following lemmas. 

\begin{lemma}\label{lem:contraction_seq}
    Let $x_0, x_1, \ldots, x_{L-1}$ be the sequence of length $L = \lceil 3 \log \log \log n \rceil + 1$ where $x_0 = 100$ and $x_i = 100^{1.5^i - 1.5^{i-1}}$ for all $1 \le i \le L - 1$. This sequence satisfies the following conditions: $x_i \geq 2$, $\prod_{i = 0}^{L-1} x_i \geq \Omega(\log n)$, and $\sum_{i = 0}^{L-1} \frac{x_i}{x_0 x_1 \ldots x_{i-1}} \leq O(1)$.
\end{lemma}
\begin{proof} $x_i \geq 2$ is trivial. Since $\prod_{i = 0}^{t} x_i = 100^{1.5^t}$ for all $0 \le t \le L-1$, and $\prod_{i = 0}^{L-1} x_i \geq 100^{1.5^{3 \log \log \log n}} = \Theta(\log n)$, the second condition follows. The third condition can be rewritten as $x_0 + \sum_{i=1}^{L-1} 100^{1.5^i - 1.5^{i-1} - 1.5^{i-1}}$. As $100^{-0.5 \cdot 1.5^i} = (\frac{1}{10})^{1.5^i} \leq (\frac{1}{10})^{0.1i}$, it is upper bounded by a constant.
\end{proof}
\begin{lemma}\label{lem:contraction_seq2}
    Let $x_0, x_1, \ldots, x_{L-1}$ be the sequence of length $L = O(\log \log \log n)$ where $x_0 = 100$ and $x_i = 100^{1.5^i - 1.5^{i-1}}$ for all $1 \le i \le L - 1$. This sequence satisfies the following conditions $x_i \geq 2$, $\prod_{i = 0}^{L-1} x_i = \Theta(\log n)$
    , and $\sum_{i = 0}^{L-1} \frac{x_i}{x_0 x_1 \ldots x_{i-1}} \leq O(1)$.
\end{lemma}
\begin{proof}
    We take the sequence of \cref{lem:contraction_seq}. Then, we truncate the suffix of $x$ and scale $x_{L-1}$ down appropriately so that $\prod_{i = 0}^{L-1} x_i$ will fit exactly in $\Theta(\log n)$. This modification does not violate any of the conditions in \cref{lem:contraction_seq}.
\end{proof}
\subsection{Batch Dynamic Maintenance of Nested Contraction}\label{sec:batchdynamic_true}
We now proceed to proving \cref{thm:maintrue}. \cref{thm:maintrue} is obtained via iterations of contractions, as outlined above. In our data structure of \cref{thm:maintrue}, we maintain each level of the nested contractions in a \textit{layer} $0, 1, \ldots, L$, where layer $i$ holds the information $G_i, H_i, f_i$ that is defined identically as in \cref{sec:contraction_nest}. For the layer $L$, the graph is maintained with \cref{thm:mainnear}. For other layers, the information is maintained using the following data structures. 

\begin{itemize}
    \item Each vertex in $V_i$ is indexed as an unique integer in range $[0, |V_i| - 1]$.
    \item For each vertex $v \in [0, |V_i| - 1]$, we store a binary search tree $\textsc{Adj}_i(v)$, which contains all the edges in $E_i$ that is incident with the vertex $v$. Each entry is assigned a random real value of range $[0, 1)$, and is ordered in ascending order of the random value. 
    \item For each vertex $v \in [0, |V_i| - 1]$, we store an index $\textsc{Head}_i(v)$ of range $[-1, |V_{i+1}| - 1]$, denoting the function value $f_i(v)$ according to the index of $f_{i}(v)$ in $V_{i+1}$. If $\textsc{Head}_i(v) = -1$, then $f_i(v) = \bot$. 
    \item A hash table $H_i$, that stores all edges of either of the two types: $(u, v) \in E_i$ such that either $\textsc{Head}_i(u) = -1$ or $\textsc{Head}_i(v) = -1$, or an edge between $\textsc{Head}_i(v)$ and $v$, for $\textsc{Head}_i(v) \neq v, \textsc{Head}_i(v) \neq -1$. This hash table contains exactly the edges of $H_i$ as in \cref{sec:contraction_nest}.
    \item A hash table $\textsc{NextLevelEdges}_i$, stores a tuple of edges \\$(\textsc{Head}_i(u), \textsc{Head}_i(v), e)$ for all $e = (u, v) \in E_i$ which satisfies the following: $\textsc{Head}_i(u) \neq -1, \textsc{Head}_i(v) \neq -1$, and $\textsc{Head}_i(u) \neq \textsc{Head}_i(v)$.
    \item A hash table $\textsc{BwdCorrespondence}_i$, stores a mapping from the edge $e^\prime \in E_{i+1}$ to the edge $e \in E_i$, where $e$ is an edge \textit{corresponding} to $e^\prime$. The mapping exists for all the edges of $e^\prime \in E_{i+1}$. The corresponding edges of $e^\prime$ are chosen arbitrarily.
    \item A hash table $\textsc{FwdCorrespondence}_i$, stores an inverse mapping of $\textsc{BwdCorrespondence}_i$. 
\end{itemize}

We assume that there is no two random values in $\textsc{Adj}_i(v)$ that are equal, which can be assumed with high probability if the random value has sufficiently large bits (such as $1000 \log n)$ and if there are at most polynomially many updates to the data structure. If the number of updates is superpolynomial, we can reinitialize the data structure for every $O(n^3)$ iteration without affecting our asymptotic bounds.

We first discuss the additional invariant we enforce on $G_i, H_i, f_i$ in our fully dynamic algorithm that will be maintained throughout the run. Let $G_i = (V_i, E_i)$. We first assume $V_i$ to be fixed throughout the entire run: In our data structure, the sampling of vertices is independent of the graph's edges. The sampling of vertices is also independent of the updates given to the data structure by our oblivious adversary assumption. Hence, we compute the set $V_i$ at the initialization stage of our algorithm, along with its corresponding index of range $[0, |V_i|-1]$. For the function $f_i$, we choose it in the following way:
\begin{itemize}
    \item If $v$ has a neighbor $w$ that is in $V_{i+1}$, take the $w$ that \textit{minimizes} the random value in $\textsc{Adj}_i(v)$, and declare $\textsc{Head}_i(v) = w$. 
    \item Otherwise, $\textsc{Head}_i(v) = -1$.
\end{itemize}

This invariant on $f_i$ is strictly enforced and is crucial in our algorithm analysis.

To initialize the data structure, we compute all $V_i$ randomly, as described earlier. Beginning from the layer $0$, we populate a binary search tree $\textsc{Adj}_i(v)$ and compute the corresponding index $\textsc{Head}_i(v)$ in $O(\log n)$ depth and $O(m \log n)$ work. From that information, we can initialize all the hash tables and, consequently, the set of edges $E_{i+1}$. Repeating this for layer $1, 2, \ldots, L-1$, we can initialize all but layer $L$ in $O(\log^2 n)$ depth and $O(m \log n)$ work. The final layer, layer $L$, can be initialized with \cref{thm:mainnear} with $k = O(\log n)$, which takes $O(\log^2 n)$ depth and $O(m \log n)$ work. 

Now, we discuss the update of our data structure. We define the process $\textsc{Update}(i, U_{ins}, U_{del})$, which deletes all the edges $U_{del}$ from the $G_i$, and inserts all the edges $U_{ins}$ to the $G_i$. In this way, the update to our data structure can be done by either invoking $\textsc{Update}(0, U, \emptyset)$ or $\textsc{Update}(0, \emptyset, U)$, depending on if $U$ is an insertion or deletion updates. 

An $\textsc{Update}(i, U_{ins}, U_{del})$ first processes the deletion of edges in $U_{del}$ and then processes the insertion of edges in $U_{ins}$. If $i = L$, these can be done using the data structure of \cref{thm:mainnear}. Otherwise, we will describe an algorithm that will maintain all the data structures and return two sets of edges $\textsc{Next}_{ins}, \textsc{Next}_{del}$, that should be deleted in the layer $i+1$.

We first demonstrate the algorithm to delete the edges in $U_{del}$. For each edge $(u, v) \in U_{del}$, they belong to either one of the four cases:
\begin{itemize}
    \item \textbf{Case D1.} $\textsc{Head}_i(u) = -1$ or $\textsc{Head}_i(v) = -1$.
    \item \textbf{Case D2.} $\textsc{Head}_i(u) \neq \textsc{Head}_i(v)$
    \item \textbf{Case D3.} $\textsc{Head}_i(u) = \textsc{Head}_i(v)$, and $\textsc{Head}_i(u) \notin \{u, v\}$
    \item \textbf{Case D4.} $\textsc{Head}_i(u) = \textsc{Head}_i(v)$, and $\textsc{Head}_i(u) \in \{u, v\}$
\end{itemize}

It can be determined in $O(1)$ work per edge and $O(1)$ depth to check which case each edge belong to. Then, we process each of these cases in the order given below:

\paragraph{Processing Case D1 Edges} We remove each edge from $\textsc{Adj}_i$, and $H_i$. There is no other data structure to be updated. Specifically, we do not need to change $\textsc{Head}_i$. No work is necessary for layer $i+1$ and beyond. These take $O(\log n)$ work per edge, and $O(\log n)$ depth.

\paragraph{Processing Case D2 Edges} We remove each edge from $\textsc{Adj}_i$ and $\textsc{NextLevelEdges}_i$. We do not change $\textsc{Head}_i$. However, we might need to change $\textsc{BwdCorrespondence}_i$ and $\textsc{FwdCorrespondence}_i$, since the removed edge could be a corresponding edge for some $e^\prime \in E_{i+1}$. If we can find a replacement for the corresponding edge of $e^\prime$ from $\textsc{NextLevelEdges}_i$, we assign it and finish the work. Otherwise, we remove the entries and add $e^\prime$ into $\textsc{Next}_{del}$ to delete these in layer $i+1$. These take $O(\log n)$ work per edge. Everything can be done in parallel, so this requires $O(\log n)$ depth.

\paragraph{Processing Case D3 Edges} These are the edges with $\textsc{Head}_i(u) = \textsc{Head}_i(v)$ that are not a part of $H_i$. We remove each edge from $\textsc{Adj}_i$. There is no other data structure to be updated. Specifically, we do not need to change $\textsc{Head}_i$. No work is necessary for layer $i+1$ and beyond. These take $O(\log n)$ work per edge. Everything can be done in parallel, so this requires $O(\log n)$ depth.

\paragraph{Processing Case D4 Edges} These are the edges with $\textsc{Head}_i(u) = \textsc{Head}_i(v)$ in $H_i$. Without loss of generality, assume that $\textsc{Head}_i(u) = u$. We remove each edge from $\textsc{Adj}_i$ and $H_i$. Then, we recalculate the $\textsc{Head}_i(v)$ for all endpoints of $v$. For that, we iterate the $\textsc{Adj}_i$, in parallel, to find the new value of $\textsc{Head}_i(v)$. This will change the value $\textsc{Head}_i(v)$, possibly to another vertex or $-1$. If $\textsc{Head}_i(v) = -1$, we add all edges of $\textsc{Adj}_i(v)$ into the hash table $H_i$; otherwise, each edge will start to connect different vertices of $V_{i+1}$. We apply these changes in $\textsc{NextLevelEdges}_i$, $\textsc{BwdCorrespondence}_i$, and $\textsc{FwdCorrespondence}_i$. If such an update introduces or removes the edge in $E_{i+1}$, we add those edges into $\textsc{Next}_{ins}$ or $\textsc{Next}_{del}$ and let these be deleted in layer $i+1$. Everything can be done in parallel (specifically, the procedure of finding the new $\textsc{Head}_i(u)$ is independent of other edges), so this requires $O(\log n)$ depth. For each edge, we do $O(\deg(v) \log n)$ work to apply the changes, and we add at most $\deg(v)$ edges in $\textsc{Next}_{ins}$ or $\textsc{Next}_{del}$.

The suggested upper bound of $O(\deg(v) \log n)$ work and $\deg(v)$ new edges in $\textsc{Next}_{ins}$ or $\textsc{Next}_{del}$ is prohibitively large. However, the expected work per edge is actually at most $O(\log n)$, and we will only create $O(1)$ edges in expectation. We defer the proof of this fact until after our complete demonstration of the algorithm.

Next, we demonstrate the algorithm to insert the edges in $U_{ins}$. For each edge $(u, v) \in U_{ins}$, they belong to either one of the six cases:

\begin{itemize}
    \item \textbf{Case I1.} $\textsc{Head}_i(u) \neq u$, $\textsc{Head}_i(v) \neq v$, and either $\textsc{Head}_i(u) = -1$ or  $\textsc{Head}_i(v) = -1$.
    \item \textbf{Case I2.} $\textsc{Head}_i(u) \neq u$, $\textsc{Head}_i(v) \neq v$, $\textsc{Head}_i(u) \neq -1$, and $\textsc{Head}_i(v) \neq -1$.
    \item \textbf{Case I3.} $\textsc{Head}_i(u) = u$, and $\textsc{Head}_i(v) = v$.
    \item \textbf{Case I4.} Exactly one of $\textsc{Head}_i(u) = u$, and $\textsc{Head}_i(v) = v$ holds. And, either $\textsc{Head}_i(u) = -1$ or $\textsc{Head}_i(v) = -1$.
    \item \textbf{Case I5.} Exactly one of $\textsc{Head}_i(u) = u$, and $\textsc{Head}_i(v) = v$ holds. And, neither $\textsc{Head}_i(u) = -1$ nor $\textsc{Head}_i(v) = -1$.
\end{itemize}

It can be determined in $O(1)$ work per edge and $O(1)$ depth to check which case each edge belong to. Then, we process each of these cases in the order given below:

\paragraph{Processing Case I1 Edges} We insert each edge to $\textsc{Adj}_i$, and $H_i$. There is no other data structure to be updated. No work is necessary for layer $i+1$ and beyond. These take $O(\log n)$ work per edge. Everything can be done in parallel, so this requires $O(\log n)$ depth.

\paragraph{Processing Case I2 and I3 Edges} For I2 edges, we insert each edge to $\textsc{Adj}_i$. We do not need to change $\textsc{Head}_i$. If $\textsc{Head}_i(u) = \textsc{Head}_i(v)$, we are done. Otherwise, we insert the edges in $\textsc{NextLevelEdges}_i$. If the added edge creates a new edge in $E_{i+1}$, we update $\textsc{BwdCorrespondence}_i$ and $\textsc{FwdCorrespondence}_i$, and put the edge that need to be created in $E_{i+1}$ in $\textsc{Next}_{ins}$. These take $O(\log n)$ work per edge. Everything can be done in parallel, so this requires $O(\log n)$ depth. The process for I3 Edges is identical.

\paragraph{Processing Case I4 Edges} We insert each edge to $\textsc{Adj}_i$. WLOG, assume that $\textsc{Head}_i(u) = u$. For each vertex $v$, we need to recalculate $\textsc{Head}_i(v)$ and move all the edges incident to $v$ in the appropriate data structures. In other words, we remove all of them from $H_i$, and add back to either $H_i$, or ($\textsc{NextLevelEdges}_i$, $\textsc{BwdCorrespondence}_i$ and $\textsc{FwdCorrespondence}_i$), or nowhere, depending on the $\textsc{Head}_i$ value of the opposite endpoint. Let $\deg(v)$ be the degree of the vertex $v$, after the insertion is processed. For each edge, we need $O(\deg(v) \log n)$ work to apply the changes, and we may add at most $\deg(v)$ edges in $\textsc{Next}_{ins}$. This also seems to be prohibitive, but it turns out that expected work per edge here is at most $O(\log n)$, and $O(1)$ edges are added to $\textsc{Next}_{ins}$ in expectation, as in Case D4. Again, we defer this argument to the analysis.

\paragraph{Processing Case I5 Edges} We insert each edge to $\textsc{Adj}_i$. WLOG, assume that $\textsc{Head}_i(u) = u$. For each of the vertex $v$, we may need to recalculate $\textsc{Head}_i(v)$, since our invariant forces us to have $\textsc{Head}_i(v)$ as the one that minimizes the random value in $\textsc{Adj}_i(v)$, among the ones with $\textsc{Head}_i(u) = u$. By comparing the random value between the edges directed to $\textsc{Head}_i(v) = u$ in $\textsc{Adj}_i(v)$ and the newly inserted edge, we can check whether the recalculation is necessary. If it is, we recalculate the $\textsc{Head}_i(u)$ and update all $\deg(v)$ incident edges to reflect the changes, identically as done in Case D4. This also suggests the upper bound of $O(\deg(v) \log n)$ work and $\deg(v)$ new edges in $\textsc{Next}_{ins}$ or $\textsc{Next}_{del}$ that is prohibitively large, but we will later show that the expected work per each edge here is actually at most $O(\log n)$, and we will only create $O(1)$ edges in expectation. Again, we defer this argument to the analysis.

Finally, we can call $\textsc{Update}(i+1, \textsc{Next}_{ins}, \textsc{Next}_{del})$ to update the layer $i+1$, which will recursively reflect the changes all the way to the layer $L$. Eventually, $\textsc{Update}(i+1, \textsc{Next}_{ins}, \textsc{Next}_{del})$ will return a set of edge updates $\delta H_{ins}, \delta H_{del}$. We map those edges in $G_{i+1}$ to the edges in $G_i$ by an access to $\textsc{BwdCorrespondence}_i$ and return them with the updates that had been applied to the current layer $H_i$. This finishes the description of our algorithm.

Note that the decision of our randomized algorithm is completely independent of the returned result of the deeper layer, or the algorithm of \cref{thm:maintrue}. The only operation we do with the returned value is to apply the mapping $\textsc{BwdCorrespondence}_i$. Hence, an assumption of the oblivious adversary at layer $i$ can be safely assumed in layer $i+1$ and so forth.

We begin the analysis by revisiting the claim left unproven in Case D4, Case I4, and Case I5. In all cases, our algorithm requires $O(1)$ updates of edges in all circumstances, with one exception: if an update changes the value $\textsc{Head}_i(v)$, the algorithm iterates all $\deg(v)$ edges in the data structure. 

For each element (an edge) $e = (v, w) \in \textsc{Adj}_i(v)$, we define the tuple $(unmark_e, rand_e)$ as the following: $unmark_e$ is true if $w \notin V_{i+1}$, and $rand_e$ is a random value in $[0, 1)$ assigned when we insert an element in $\textsc{Adj}_i(v)$. All $(unmark_e, rand_e)$ are independent and drawn from the same distribution: $unmark_e$ depends on whether the vertex $w$ is sampled from the set $V_i$, which is drawn uniformly with probability $\frac{1}{x}$. Also, we can assume that the randomness is fixed before the update begins, removing the possible dependency between the adversary and our random variables.

In all of our Cases D4, I4, and I5, we needed $O(\deg(v) \log n)$ extra work and $\deg(v)$ edges to insert in $\textsc{Next}_{ins}, \textsc{Next}_{del}$ if the value $\textsc{Head}_i(v)$ changed after the update. The value $\textsc{Head}_i(v)$ is defined by the minimum possible $rand_e$ among all element $e \in \textsc{Adj}_i(v)$ with $unmark_e = 0$. Hence, for all cases, the recalculation of $\textsc{Head}_i(v)$ happens only if the minimum value of $(unmark_e, rand_e)$ (in lexicographical order) has changed. Every element of $(unmark_e, rand_e)$ is drawn from the same distribution. Thus, the insertion of a single element in a data structure of size $n$ changes the minimum with probability $\frac{1}{n + 1}$, and the deletion of a single element in a data structure of size $n$ changes the minimum with probability $\frac{1}{n}$, regardless of the choice of the element adversary makes. Hence, if $\textsc{Adj}_i(v)$ had only a single element inserted or deleted, our expected work bound is $O(\log n)$, and we only need $1$ edge to insert in $\textsc{Next}_{ins}, \textsc{Next}_{del}$. Although our parallel algorithm inserts/deletes multiple elements simultaneously, the same analysis applies.

We review the work, depth, and recourse bounds (i.e., the amortized size of $|\delta H_{ins}| + |\delta H_{del}|$). As shown earlier, the initialization stage takes $O(\log^2 n)$ depth and $O(m \log n)$ work. For each edge update in layer $i$, the expected number of new edge updates it makes in layer $i+1$ is $O(1)$. Since each layer is independent, the total expected number of new edge updates is $O(1)^{\log \log \log n} = \text{poly}(\log \log n)$. In our data structure, each edge update in the layer takes $O(\log n)$ expected time. 

Each edge update in \cref{thm:mainnear} takes  $O(\log^3 n)$ amortized work per edge update. Hence, each inserted and deleted edge takes $O(\log^3 n \cdot \text{poly}(\log \log n))$ expected work. Every level takes $O(\log n)$ worst-case depth to process, so the depth term is dominated by $O(\log^3 n)$ worst-case depth of \cref{thm:mainnear}. After each layer, the number of edges to be updated in the next layer is multiplied by a constant. This amounts to $O(1)^{\log \log \log n} = \text{poly}(\log \log n)$ edge update, which is again dominated by \cref{thm:mainnear}.


\section{Ultra-Sparse Spanners}

Here, we present our batch-dynamic ultra-sparse spanner with sublinear extra number of edges, achieving via a single contraction, and thus prove \cref{thm:mainultra}.

\subsection{The Contraction Procedure}\label{sec:contraction_ultra}
We describe the procedure $\textsc{ContractUltra}(G, x)$, which takes a simple graph $G = (V, E)$ with $n$ vertices and an integer parameter $x$, and returns a subproblem of size $O(\frac{|V|}{x})$. Formally, we state the following:

\begin{lemma}\label{lem:contraction-ultra}
    There is a procedure $\textsc{ContractUltra}(G, x)$ which takes a simple graph $G = (V, E)$ with $n$ vertices and a parameter $2 \le x \le O(\log^2 n)$, and returns a tuple $(G^\prime, H, f)$ such that the following holds with high probability:
\begin{itemize}
    \item $G^\prime = (V^\prime, E^\prime)$ is a graph, $f : V \rightarrow V^\prime \cup \{\bot \}$ is a function, $H \subseteq E$ is a subset of edges.
    \item $V^\prime$ is not empty.
    \item For any $y \in V^\prime$, $f(y) = y$ holds.
    \item $E[|V^\prime|] \leq \frac{2n}{x}$.
    \item $|H| \leq n - 1$.
    \item For any $L$-spanner of $G^\prime$ denoted as $H^\prime$, we can find a $21x\log x \cdot (L+1)$-spanner of size $|H^\prime| + |H|$, that contains all the edges of $H$.
\end{itemize}
\end{lemma}

In this section, we will describe the simple sequential version of $\textsc{ContractUltra}(G, x)$ to make the proof more clear. We will later describe ingredients to adapt the $\textsc{ContractUltra}(G, x)$ into a parallel batch-dynamic model that will be used to prove \cref{thm:mainultra}.

We begin by computing the set of vertices $D \subseteq V$, uniformly sampled from $V$ with probability $\frac{1}{x}$. For each vertex $v \in V$, we classify them as \textit{heavy} if $\deg(v) \geq 10 x \log x$, and \textit{light} otherwise. Here, $\deg(v)$ is the degree of vertex $v$ in $G$. Finally, let $P$ be an arbitrary permutation of $V$ used for the tie-breaking.

We now present the definition of function $f$, which indicates the contracted cluster to which every vertex belongs. For each \textit{heavy} vertex $v$, the function $f$ is defined in the following way.
\begin{itemize}
    \item If there is a vertex $w$ in $D$ such that the distance from $v$ to $w$ is at most $1$, take such $w$ and declare $f(v) = w$. If there is more than one such vertex, take the one that is the closest. If there is still more than one such vertex, take the one that occurs earliest at the permutation $P$. We call such vertex to be \textit{clustered}.
    \item Else, $f(v) = v$, and we call such vertex to be \textit{unclustered}.
\end{itemize}

Let $D^\prime$ be the set of unclustered heavy vertex $v$. For the \textit{light} vertex $v$, the function $f$ is defined in a slightly different way:
\begin{itemize}
    \item If there is a vertex $w$ in $D \cup D^\prime$ such that the distance from $v$ to $w$ is at most $10x \log x$, take such $w$ and declare $f(v) = w$. If there is more than one such vertex, take the one that minimizes the distance. If there is still more than one such vertex, take the one that occurs earliest at the permutation $P$.
    \item Else, if $v$ belongs to a component with at most $10x \log x$ vertices, declare $f(v) = \bot$.
    \item Else, $f(v) = v$.
\end{itemize}

We want to set $H$ as the spanning tree of each cluster with a short diameter. To show that this is possible, we need to show that our clustering satisfies several desirable properties:

\begin{lemma}\label{lem:wellformed}
    For any $v \in V^\prime$, $f(v) = v$.
\end{lemma}
\begin{proof}
    For any vertex $v$ where $f(v) \neq v, f(v) \neq \bot$, we have $f(v) \in D \cup D^\prime$, where $f(f(v)) = f(v)$ holds by definition.
\end{proof}

Next, we want to show that each cluster forms a connected subgraph with low diameter.

\begin{lemma}\label{lem:lowdiamarea}
    For any vertex $w \in V^\prime$, let $S_w = \{v | f(v) = w, v \in V\}$. In the induced subgraph of $S_w$, any vertex in $S_w$ can be reached from $w$ with a path of length at most $10x \log x$.
\end{lemma}
\begin{proof}
    By definition, the distance between $w$ and any vertex in $S_w$ is at most $10x \log x$ in $G$. Hence, it suffices to prove that, for any vertex $z \in S_w$, a shortest path between $w$ and $z$ can not contain any vertex outside of $S_w$. Suppose not. Let $P$ be a shortest path between $w$ and $z$, and let $y \in P$ be a vertex in $P$ such that $f(y) \neq w$. Observe:

    \begin{itemize}
        \item $z$ is a light vertex, since $dist(w, z) \geq 2$.
        \item $w \in D \cup D^\prime$, as otherwise there can not be a vertex $z \neq w$ such that $f(z) = w$.
        \item $f(y) \in D \cup D^\prime$. Suppose not, then $y$ should be the light vertex with no vertex of $D \cup D^\prime$ within distance $10x \log x$, which is not true due to $w$.
    \end{itemize}
    
    Finally, note that for any vertex $a$, if $dist(y, a) < dist(y, w)$, then $dist(z, a) < dist(z, w)$, since 
\begin{align*}
dist(y,a) < dist(y,w) \quad &\\
dist(z,y) + dist(y,a) < dist(z,y) + dist(y,w) \quad &\\
dist(z,a) < dist(z,y) + dist(y,w) \quad & (\text{triangle inequality})\\
dist(z,a) < dist(z,w) \quad & (\text{$y$ in a shortest path between $w$ and $z$})
\end{align*}
and for the same reason, if $dist(y, a) \leq dist(y, w)$, then $dist(z, a) \leq dist(z, w)$.

Suppose that $f(y) = y$. Recall that $z$ is light, and $y \in D \cup D^\prime$. As $dist(y, z) < dist(w, z)$, $f(z) \neq w$, a contradiction. Hence, $f(y) \neq y, f(y) \neq \bot$. We consider two cases.
\begin{itemize}
    \item $y$ is heavy: There is a vertex $f(y) \in D$ where $1 = dist(y, f(y)) \leq dist(y, w)$ and $f(y)$ occurs earlier than $w$ in the permutation $P$. Consequently, we have $dist(z, f(y)) \leq dist(z, w)$, and $f(y)$ occurs earlier than $w$ in the permutation $P$. As such, $f(z) \neq w$ holds, a contradiction.
    \item $y$ is light: We have $f(y) \in D \cup D^\prime$, where either $dist(y, f(y)) < dist(y, w)$ or $dist(y, f(y)) = dist(y, w)$ and $f(y)$ occurs earlier than $w$ in the permutation $P$. This implies that either $dist(z, f(y)) < dist(z, w)$ or $dist(z, f(y)) \leq dist(z, w)$ and $f(y)$ occurs earlier than $w$ in the permutation $P$. Consequently, we have $f(z) \neq w$, a contradiction.
\end{itemize}
\end{proof}

Now, we declare $H$ as the union of the following:
\begin{itemize}
    \item A collection of shortest path tree rooted at $w$, for each induced subgraph of $S_w$ for $w \in V^\prime$, which exists by \cref{lem:lowdiamarea}.
    \item Any spanning forest of the subgraph of $G$ induced by vertices with $f(v) = \bot$. 
\end{itemize}

Finally, the graph $G^\prime = (V^\prime, E^\prime)$ is constructed by removing all vertices with $f(v) = \bot$, contracting each pair of vertices $(v, f(v))$, and remove all self-loops or duplicate edges to make it a simple graph. Formally, we set $V^\prime = \{f(v) | v \in V, f(v) \neq \bot\}$, and for all pair of $(u^\prime, v^\prime)$ such that $u^\prime \neq v^\prime, \{u^\prime, v^\prime\} \in V^\prime$, we put and edge between $u^\prime$ and $v^\prime$ in $E^\prime$ if there exists an edge $(u, v)$ in $E$ such that $f(u) = u^\prime, f(v) = v^\prime$. 

This finishes the description of the procedure $\textsc{ContractUltra}(G, x)$. We check that our implementation satisfies all statements of \cref{lem:contraction-ultra}.

\begin{proof}[Proof of \cref{lem:contraction-ultra}]
    By definition, $G^\prime, f, H$ are objects of a given type. By \cref{lem:wellformed}, $f(y) = y$ holds for all $y \in V^\prime$. The probability of $D$ being empty is $\exp(-\Omega(\frac{n}{x}))$, which is negligible under our parameters. As $D \subseteq V^\prime$, the probability of $V^\prime$ being empty is also negligible. $H$ is a forest, and consequently $|H| \leq n - 1$. 
    
    We prove that the expected number of vertices in $V^\prime$ is at most $\frac{2n}{x}$. By \cref{lem:wellformed}, it suffices to prove that there are at most $\frac{2n}{x}$ vertices with $f(v) = v$. The expected size of $D$ is $\frac{n}{x}$. For a heavy vertex to have $f(v) = v$ while $v \notin D$, none of its neighbors should belong to $D$. This happens with probability at most $(1 - \frac{1}{x})^{10 x \log x + 1} \leq x^{-10}$, which means there are at most $\frac{n}{x^{10}}$ such vertices in expectation. For a light vertex to have $f(v) = v$ while $v \notin D$, there should be at least $10x \log x$ vertices that are reachable by $v$ within distance at most $10x \log x$. Since all of them should not belong to $D$, this happens with probability at most $(1 - \frac{1}{x})^{10x \log x + 1} \leq x^{-10}$, which means there are at most $\frac{n}{x^{10}}$ such vertices in expectation. Since $x \geq 2$, this sums to at most $\frac{1.1n}{x}$.
    
    Finally, we show an algorithm for finding a $21x \log x \cdot (L+1)$-spanner of size $|H^\prime| + |H|$ that contains all edges of $H$, given the $L$-spanner $H^\prime$. For edge $(u^\prime, v^\prime) \in E^\prime$, we pick any $(u, v) \in E$ such that $\textsc{Head}(u) = u^\prime, \textsc{Head}(v) = v^\prime$, and say $(u, v)$ is \textit{corresponding} to $(u^\prime, v^\prime)$. Clearly, every edge in $E^\prime$ can find a corresponding edge in $E$. Our spanner consists of two types of edges: One, being all the edges of $H$, and another, being the set of corresponding edges of $H^\prime$. Clearly, we have a spanner of size $|H^\prime| + |H|$ that contains all the edges of $H$. It only remains to show that the stretch is at most $21x \log x \cdot (L+1)$. 
    
    There are three types of edges $(u, v) \in E$ that are omitted in the spanner:
    \begin{itemize}
        \item edges with either $f(u) = \bot$ or $f(v) = \bot$, 
        \item edges with $f(u) = f(v)$, and
        \item edges where $(f(u), f(v)) \in E^\prime$ but is not corresponding to $(f(u), f(v))$.
    \end{itemize}

    For the first type of edges, both $u$ and $v$ should belong to the same connected component of at most $10x \log x$ vertices. Hence, a spanning forest over a vertices with $f(v) = \bot$ will preserve a path between such vertices with at most $10 x \log x$ edges.
    
    Now, we can assume $f(u) \neq \bot, f(v) \neq \bot$. For the second type of edges, let $w = f(u) = f(v)$. By \cref{lem:lowdiamarea}, one can pass through $u$ and $v$ via at most $20x \log x$ edges in $H$ by passing through $w$. For the third type of edges, as $H^\prime$ is a $L$-spanner of $G^\prime$, one can move through $f(u)$ and $f(v)$ in $G^\prime$ using at most $L$ edges in $H^\prime$. Replacing each of the edges in the path with the corresponding edges, we need to fill in between each edges with the path between two vertices with same $f(*)$ (which are guaranteed to not be $\bot$). By \cref{lem:lowdiamarea}, we need to add $10x \log x\cdot (L+1)$ edges in $H$, which is used to move between vertices that share the same $f(v)$. This amounts to $20x\log x \cdot(L+1) + L$ edges in total, proving the $21x \log x \cdot (L+1)$ upper bound on the stretch.
\end{proof}

\subsection{Batch-Dynamic Maintenance of Contraction}
We now proceed to proving \cref{thm:mainultra}. In our data structure of \cref{thm:mainultra}, we maintain the contractions by storing the information of $f, G^\prime, H$ defined as in \cref{sec:contraction_ultra}. These information are maintained using the following data structures:

\begin{itemize}
    \item Each vertex in $V$ is indexed as a unique integer in range $[0, |V| - 1]$.
    \item For each vertex $v \in [0, |V| - 1]$, we assign a random value $rand_v$ of range $[0, 1)$ represented in $1000 \log n$ bits each.
    \item For each vertex $v \in [0, |V| - 1]$, we assign an indicator $unmark_v$, which is $1$ with probability $1 - \frac{1}{x}$. Here, $v \in D$ if and only if $unmark_v= 0$.
    \item For each vertex $v \in [0, |V| - 1]$, we store a binary search tree $\textsc{Adj}(v)$, which contains all the edges in $E$ that is incident with the vertex $v$. For an edge $(v, w) \in E$, we add a tuple $(unmark_v, rand_v, v)$ into the data structure $\textsc{Adj}(w)$, and a tuple $(unmark_w, rand_w, w)$ into the data structure $\textsc{Adj}(v)$. Each entries are ordered in the ascending order of the tuple. 
    \item For each vertex $v \in [0, |V| - 1]$, we store an index $\textsc{Head}(v)$ of range $[-1, |V| - 1]$, denoting the function value $f(v)$. If $\textsc{Head}(v) = -1$, then $f(v) = \bot$. 
    \item A hash table $\textsc{NextLevelEdges}$, stores a tuple of edges $(\textsc{Head}(u), \textsc{Head}(v), e)$ for all $e = (u, v) \in E_i$ where $\textsc{Head}(u) \neq -1, \textsc{Head}(v) \neq -1, \textsc{Head}(u) \neq \textsc{Head}(v)$.
    \item A hash table $\textsc{BwdCorrespondence}$, stores a mapping from the edge $e^\prime \in E^\prime$ to the edge $e \in E$, where $e$ is an edge \textit{corresponding} to $e^\prime$. The mapping exists for all the edges of $e^\prime \in E^\prime$. The corresponding edges of $e^\prime$ are chosen arbitrarily.
    \item A hash table $\textsc{FwdCorrespondence}$, stores an inverse mapping of $\textsc{BwdCorrespondence}$. 
    \item A sparse spanner data structure $G^\prime$ over a graph $(V, E^\prime)$. $G^\prime$ is implemented with \cref{thm:maintrue}.
    \item A hash table $H_1$, that stores all edges that consists the shortest path tree for each induced subgraph $S_w$ for $w \in V^\prime$.
    \item A dynamic spanning forest data structure $H_2$, that maintains the spanning forest of edges $(u, v)$ where $\textsc{Head}(u) = \textsc{Head}(v) = \bot$. $H_2$ is implemented with \cite{acar2019parallel}.
\end{itemize}

The random value of $rand_v$ implements the permutation $P$ of \cref{sec:contraction_ultra}: Our algorithm will consider $P$ as an order of vertex sorted in the increasing order of $rand_v$. We assume that there are no two random values of $rand_v$ that are equal, which can be assumed with high probability if there are at most polynomially many updates to the data structure. If we can not assume the number of updates to be polynomially bounded, we can reinitialize the data structure every polynomial number of times, which does not affect our asymptotic work bound and depth. 

The initialization of the data structure is performed in the order of the objects listed in the previous paragraph. We assign $rand_v, unmark_v$ randomly, and initialize the BST $\textsc{Adj}(v)$. The value $\textsc{Head}(v)$ is first computed for the heavy vertices, and then for the small vertices. For each of the vertex, we compute the value $\textsc{Head}(v)$ with the procedure $\textsc{ComputeHead}(v)$, which works in the following way:
\begin{itemize}
    \item For heavy vertices, iterate through $\textsc{Adj}(v)$ in parallel.
    \item For light vertices, we perform a bounded BFS of \cref{lem:app_bfs} of depth $10 x \log x$, but do not branch on any heavy vertices. Then, we compute the $\textsc{Head}(v)$ in the following way: For the light vertices explored in the BFS, take the one with the smallest distance (tiebroken by $rand_v$). For the heavy vertices $w$, we take $\textsc{Head}(w)$ as a candidate for our consideration. If $\textsc{Head}(w)$ was visited throughout our BFS procedure, we know the distance between $\textsc{Head}(w)$ and $v$. Otherwise, we can consider the distance between $\textsc{Head}(w)$ and $v$ to be greater than the distance between $w$ and $v$ by $1$. 
\end{itemize}

In \cref{alg:computehead}, a detailed exposition of the $\textsc{ComputeHead}(v)$ procedure is demonstrated as a pseudocode. There are two non-trivial parts in this algorithm: One is that we are not branching from a heavy vertex, and another is that from that heavy vertex, we consider a specific candidate with a specific distance choice. To see why these parts are valid, note that as soon as you reach a heavy vertex $w$, there is no better candidate to be found in the area reached by passing through $w$, other than the candidate $\textsc{Head}(w)$ - This is trivial to see if $\textsc{Head}(w) = w$, and for the other case note that all possible candidates will have its distance greater or its $rand_v$ greater. For the distance part, our value of $\textsc{Dist}(w) + 1$ is an upper bound. Suppose that $\textsc{Dist}(w) \geq \textsc{Dist}(\text{Head}(w))$. Then the only reason our BFS does not reach it is that the heavy vertices block the path toward it: In that case, we will find a better candidate to $\textsc{Head}(v)$ anyway, and $\textsc{Head}(w)$ can be considered with a worser distance value.

After we compute the value $\textsc{Head}(v)$, it is straightforward to initialize $\textsc{NextLevelEdges}$, 
$\textsc{FwdCorrespondence}$, $\textsc{BwdCorrespondence}$, $G^\prime$. To compute $H_1$, each vertex $v$ where $\textsc{Head}(v) \neq -1, \textsc{Head}(v) \neq v$ should provide an edge $(w, v)$ to a vertex which $dist(\textsc{Head}(v), w) + 1 = dist(\textsc{Head}(v), v)$. For heavy vertices, this is trivial. For light vertices, we can iterate all adjacent vertices. Initializing $H_2$ is trivial.

\begin{algorithm}\caption{ $\textsc{ComputeHead}(v)$}\label{alg:computehead}
\begin{algorithmic}[1]
\small
\If{$\deg(v) \geq 10 x \log x$}
\State{Compute $\textsc{Head}(v)$ by iterating through $\textsc{Adj}(v)$ in parallel}
\State{End the procedure}
\EndIf
\State{Initialize an empty hash set $\textsc{Visited}$}
\State{Initialize an empty hash map $\textsc{Dist}$}
\State{$S(0) = \{v\}$}
\State{$\textsc{Visited} = \{v\}$}
\State{$\textsc{Dist}(v) = 0$}
\For{$i = 1, 2, \ldots, 10 x \log x$}
\For{$v \in S(i - 1)$ in parallel}
\If{$\deg(v) < 10 x \log x$}
\For{$w \in \textsc{Adj}(v)$ in parallel}
\If{$\textsc{Visited}.\textsc{Contains}(w) = \textsf{FALSE}$}
\State{$\textsc{Visited}.\textsc{Add}(w)$}
\State{$S(i).\textsc{Add}(w)$}
\State{$\textsc{Dist}(v) = i$}
\EndIf
\EndFor
\EndIf
\EndFor
\EndFor
\State $\textsc{Head}(v) = -1$
\For{$w \in \textsc{Visited}$ in parallel}
\If{$deg(w) < 10 x \log x$}
\State Update $\textsc{Head}(v)$ to $w$ if needed
\EndIf
\If{$deg(w) \geq 10 x \log x$}
\If{$\textsc{Head}(w) \in \textsc{Visited}$}
\State Update $\textsc{Head}(v)$ to $\textsc{Head}(w)$ if needed
\EndIf
\If{$\textsc{Head}(w) \notin \textsc{Visited}$}
\State Update $\textsc{Head}(v)$ to $\textsc{Head}(w)$, assuming $\textsc{Dist}(w)+1 = \textsc{Dist}(\textsc{Head}(w))$
\EndIf
\EndIf
\EndFor
\end{algorithmic}
\end{algorithm}

Now, we discuss the update of our data structure. We first focus for maintaining the correct value of $\textsc{Head}$ after the batch of edge insertion or deletion, which is the most challenging part of our algorithm. We begin by updating the BST $\textsc{Adj}$ after each edge insertion and deletion. Then, we compute the set of vertices that requires the recomputation of $\textsc{Head}$, in the following way.

Consider the vertex $v$. If $v$ is heavy, the recomputations are necessary only if the update had changed the minimum of $(unmark_w, rand_w)$ in the BST $\textsc{Adj}_v$. Hence, such $v$ is one of the endpoints of the updated edges, and we can easily figure out if it requires recomputation. Let $R$ be the set of heavy vertex deemed to require recomputation by this process. For the light vertices, the recomputation is necessary only if there is a vertex within the distance $10 x \log x$ with its set of incident edges changed, or which had been inserted or deleted to $D^\prime$. For this, we do the same bounded BFS as in \cref{alg:computehead}, with the only difference being that we do not branch if a heavy vertex is in $R$. In \cref{alg:lnr}, we show a pseudocode that returns a set of light vertices that needs a recomputation, given the set of heavy vertices $R$ that needs a recomputation. 

\begin{algorithm}\caption{ $\textsc{LightNeedRecomputation}(R)$}\label{alg:lnr}
\begin{algorithmic}[1]
\small
\State{Initialize an empty hash set $\textsc{Visited}$}
\State{$S(0) = $ (set of endpoints of updated edges)}
\State{$\textsc{Visited} = S(0)$}
\For{$i = 1, 2, \ldots, 10 x \log x$}
\For{$v \in S(i - 1)$ in parallel}
\If{$\deg(v) < 10 x \log x$ or $v \in R$}
\For{$w \in \textsc{Adj}(v)$ in parallel}
\If{$\textsc{Visited}.\textsc{Contains}(w) = \textsf{FALSE}$}
\State{$\textsc{Visited}.\textsc{Add}(w)$}
\State{$S(i).\textsc{Add}(w)$}
\EndIf
\EndFor
\EndIf
\EndFor
\EndFor
\State Remove all vertices in $\textsc{Visited}$ with degree at least $10 x \log x$
\State \Return $\textsc{Visited}$
\end{algorithmic}
\end{algorithm}

Again, the non-trivial part is to branch on the heavy vertex that does not belong to $R$. The correctness of this branching follows from \cref{alg:computehead}. If a light vertex $v$ needs a heavy vertex not in $R$ to reach any vertex in $S(0)$ within distance at most $10 x \log x$, then the process $\textsc{ComputeHead}(v)$, will visit the exactly same set of vertices, which will return the exactly same set of candidates to compute $\textsc{Head}(v)$. Finally, we invoke $\textsc{ComputeHead}$ function for all vertices in $R$ in parallel, and then invoke $\textsc{ComputeHead}$ function for all vertices in $\textsc{LightNeedRecomputation}(R)$, which concludes the recomputation of the function $\textsc{Head}$.

Updating the value of $\textsc{Head}$ for the vertices will affect the value of $G^\prime, H_1, H_2$, and we need to maintain them appropriately. For each vertex with $\textsc{Head}(v)$ changed, we do the following:
\begin{itemize}
    \item For each edges incident to $v$, we update the set $\textsc{NextLevelEdges}$ accordingly. If the update on $\textsc{NextLevelEdges}$ necessiates the update for $\textsc{BwdCorrespondence}$, $\textsc{FwdCorrespondence}$, and the spanner data structure $G^\prime$. 
    \item If $\textsc{Head}(v) \neq -1, \textsc{Head}(v) \neq v$ holds after the update, we need to update the set of edges $H_1$ that stores the intra-cluster spanning forest. For this, it suffices to find a \textit{parent vertex} $par(v)$, where $dist(\textsc{Head}(v), par(v)) + 1 = dist(\textsc{Head}(v), v)$ and $\textsc{Head}(par(v)) = \textsc{Head}(v)$. For the heavy vertices, $par(v) = \textsc{Head}(v)$. For the light vertices, we can simply locate such a vertex during the $\textsc{ComputeHead}(v)$ process, where we compute the shortest path between $v$ and $\textsc{Head}(v)$. Note that, correspondingly, we need to remove the set of edges $(par(v), v)$ before we update the value $\textsc{Head}(v)$. 
    \item If $\textsc{Head}(v) = -1$, we add all incident edges into the dynamic spanning forest algorithm, to maintain the changes into $H_2$. Correspondingly, we need to remove the set of edges in $H_2$ if the vertex had $\textsc{Head}(v) = -1$ before the update.
\end{itemize}

Finally, we return the set of updated edges from the differences in $G^\prime, H_1, H_2$. This finishes the description of our algorithm. Note that the decision of our randomized algorithm is completely independent of the returned result of $G^\prime$. Hence, an adversary of $G^\prime$ can be considered oblivious.

Now, we move on to the analysis of our algorithm. The crucial part of our proof is to show that the heavy vertex of degree $\deg(v)$ have at most $\frac{1}{\deg(v)}$ probability of changing the value $\textsc{Head}(v)$. The proof of this statement is identical to the proof of \cref{thm:maintrue}, specifically in the final part of \cref{sec:batchdynamic_true}. The recalculation of $\textsc{Head}(v)$ happens only if the minimum value of $(unmark_v, rand_v)$ (in lexicographical order) has changed. Every element of $(unmark_v, rand_v)$ is drawn independently from the same distribution. Thus, the insertion of a single element in a data structure of size $n$ changes the minimum with probability $\frac{1}{n+1}$, and the deletion of a single element in a data structure of size $n$ changes the minimum with probability $\frac{1}{n}$, regardless of the choice of the element adversary makes. 

We first review that our invocation of $G^\prime$ using \cref{thm:maintrue}, will return the spanner of $O(n/x)$ edges in expectation. This is not a trivial argument, due to the way we are using the spanner: the vertex set of $G^\prime$ is defined to be $V$, which is a size-$n$ set, and under the statement of \cref{thm:maintrue}, $G^\prime$ is expected to return $O(n)$ edges instead of $n$. What we can guarantee, however, is that the expected number of non-isolated vertices (vertex with degree at least $1$) in $G^\prime$ is $O(n/x)$, by \cref{lem:contraction-ultra} and the fact that we do not add any edges that are not a part of $V^\prime$. To resolve this, we need a small white-box style modification to \cref{thm:maintrue}. Observe that, in \cref{thm:maintrue}, only non-isolated vertex can contribute to the spanner structure $H$ - $O(1)$ per each non-isolated vertex. Then, we modify the compression rate of \cref{thm:maintrue} from $x_i$ to $x_i^2$, which results in the resulting compressed graph of $O(\frac{n}{\log^2 n})$ vertices, and $O(\frac{n}{\log n})$ edges from \cref{thm:mainnear}, that is enough for our purposes. As a result, we can guarantee that our invocation of $G^\prime$ will return $O(n/x) + O(\frac{n}{\log n})$ edges in expectation, satisfying our guarantees.

Define $\tau(x) = (10 x \log x)^{x \log x}$ - for our choice of $x$, we have $\tau(x) = o(\log n)$. The initialization of the data structure initializes all the components of the data structure, and is dominated by the initialization of $G^\prime$ in \cref{thm:maintrue}. 

We revisit the update of our data structure one by one. Define $\tau(x) = (10 x \log x)^{x \log x}$, which we will assume $\tau(x) = o(\log n)$. Let $U$ be the set of vertices, whose set of incident edges is changed upon the updates. We first need to compute the set $R$, a set of heavy vertices requiring recomputation. As we've established earlier, each heavy vertex $v$ may belong to the set $R$ with probability at most $\frac{1}{\deg(v)}$. Then, we invoke the function $\textsc{LightNeedRecomputation}(R)$. In that function, we have $R \subseteq S(0)$, so the condition of $v \in R$ is only invoked in iteration $i = 1$. Hence, the total number of light vertices returned by the function $\textsc{LightNeedRecomputation}(R)$ is at most $|U| \cdot \tau(x) + \sum_{v \in R}(\frac{1}{\deg(v)} \deg(v) \tau(x)) = O(|U| \cdot \tau(x))$. The total work is linear to the size of the output, while the depth may grow up to $O(\log n)$ due to the parallel iteration of each vertices in $U$. We then run the $\textsc{ComputeHead}(v)$ procedure for each vertex, which takes $O(\log n)$ depth and $O(\deg(v))$ work for each heavy vertex and $O(\tau(x))$ depth and work for each light vertex. However, for heavy vertices we need nontrivial work for a probability at most $\frac{1}{\deg(v)}$ work, resulting in $O(\log n)$ depth and $O(\tau(x))$ work for all of each vertices. This sums to $O(\log n)$ depth and $O(|U| \cdot \tau(x)^2)$ total work. The remaining step all takes $O(\deg(v))$ update to each vertices of $R \cup \textsc{LightNeedRecomputation}(R)$. For heavy vertices, this is $O(1)$ expected update to the data structure per edges, since there is $O(\deg(v))$ updates with $\frac{1}{\deg(v)}$ probability. For light vertices, this is $O(\tau(x))$ update to the data structure per edges. 

The depth is dominated by \cite{acar2019parallel} and \cref{thm:maintrue} which requires $O(\log^3 n)$ depth. The update time follows from \cref{thm:maintrue}, which is $O(\log^3 n \cdot \text{poly}(\log \log n) \cdot \tau(x))$ in our case. The recourse of our data structure is dependent upon $G^\prime, H_1, H_2$ - for $G^\prime$ and $H_1$, we have an extra $\tau(x)$ factor for the recourse, and for $H_2$, we have at most $1$ recourse per update. This concludes the proof on our work, depth, and the amortized size of $|\delta H_{ins}| + |\delta H_{del}|$, which in turn concludes the proof of \cref{thm:mainultra}.

\section{Spanner Bundles, and Spectral Sparsifiers}
\label{sec:bundlesANDsparsifier}
Here we present our batch-dynamic parallel algorithms for spanner bundles and sparsifiers. After reviewing the definitions, we show a parallel batch-dynamic data structure to maintain a $t$-bundle spanner under decremental updates. Then, we use this algorithm to obtain a fully-dynamic work-efficient batch-dynamic data structure for spectral sparsifiers. We note that the spectral sparsifiers generalize cut sparsifiers.

\subsection{Definitions}

\paragraph{Spanner Bundles} For an undirected graph $G = (V, E)$, a \textit{$t$-bundle spanner} is a subset of edges $B = H_1 \cup H_2 \cup \ldots \cup H_t$, where each $H_i$ is a $O(\log n)$-spanner of $G \setminus (H_1 \cup \ldots \cup H_{i-1})$. In other words, $H_1$ is a spanner of $G$, $H_2$ is a spanner of $G \setminus H_1$, and so on.

\paragraph{Spectral Sparsifier, Cut Sparsifier}
We next review the definitions of sparsifiers.

\begin{definition}
Given an undirected graph $G$ with a vertex set $\{v_1, v_2, \ldots, v_n\}$, the Laplacian matrix \(\mathcal{L}_G\) is the \(n \times n\) matrix that, in row \(i\) and column \(j\), contains the negated weight \(-w_G(v_i, v_j)\) of the edge \((v_i, v_j)\). In the \(i\)-th diagonal entry, it contains the weighted degree of vertex \(v_i\), $\sum_{j=1}^{n} w_G(v_i, v_j)$. Note that Laplacian matrices are symmetric.
\end{definition}

\begin{definition}
    A \((1 \pm \varepsilon)\)-spectral sparsifier \(H\) of a graph \(G\) is a subgraph of \(G\) with weights \(w_H\) such that for every vector \(x \in \mathbb{R}^n\)
\[
(1 - \varepsilon) \, x^T L_H x \le x^T L_G x \le (1 + \varepsilon) \, x^T L_H x.
\]
\end{definition}

\begin{definition}
A \((1 \pm \varepsilon)\)-\emph{cut sparsifier} \(H\) of a graph \(G\) is a subgraph of \(G\) with weights \(w_H\) such that for every subset \(U \subseteq V\)
\[
(1 - \varepsilon) \, w_H\bigl(\partial_H(U)\bigr) \le w_G\bigl(\partial_G(U)\bigr) \le (1 + \varepsilon) \, w_H\bigl(\partial_H(U)\bigr).
\]
where $\partial_G(U)$ is a total weight of a cut $U$ in graph $G$.
\end{definition} 

Note that every \((1 \pm \varepsilon)\)-spectral sparsifier is a \((1 \pm \varepsilon)\)-cut sparsifier as well, due to the following reason: For any cut $C \subseteq V$, consider an indicator vector $x_C$ where $x_C(v)$ is $1$ if $v \in C$, and $0$ otherwise. Here, one can observe that $x_C^T \mathcal{L}_G x_C = \partial_G(C)$.

\subsection{Decremental Spanners with Monotonicity}
In this section, we prove the following lemma:

\begin{lemma}\label{lem:decr-bundle1}
    There is a randomized parallel batch-dynamic decremental data structure which, given an unweighted graph $G = (V, E)$ where $|V| = n, |E| = m$, maintains a $O(\log n)$-spanner of $O(n \log n)$ edges. Specifically, the algorithm supports the following interfaces:

    \begin{itemize}
        \item After the initialization, the algorithm returns a set of edges forming an $O(\log n)$-spanner of the given graph, which has $O(n \log n)$ edges.
        \item After each edge deletion updates, the algorithm returns a pair of edge sets $(\delta H_{ins}, \delta H_{del})$, representing the set of edges that are newly inserted or deleted into the $O(\log n)$-spanner. The expected total number of edges in $\delta H_{ins} \cup \delta H_{del}$ throughout all updates will be at most $O(n \log^3 n)$.
    \end{itemize}

    The algorithm takes:
\begin{itemize}
    \item for initialization, $O(m \log^2 n)$ work, and $O(\log^2 n)$ worst-case depth,
    \item for any batch of edge deletions, $O(\log^3 n)$ expected amortized work per deleted edge, and $O(\log^3 n)$ worst-case depth for the entire batch.
\end{itemize}
The above statements hold with high probability against an oblivious adversary.
\end{lemma}

The main difference between the statement of \cref{lem:mainnear} and \cref{lem:decr-bundle1} is that we have an upper bound on the total number of edges in $\delta H_{ins}$ and $\delta H_{del}$. Equivalently, this means that the number of edges that will ever be in the spanner in the entire run of \cref{lem:decr-bundle1} is bounded to $O(n \log^3 n)$, independent of the initial number of edges. Compared to \cref{lem:mainnear}, the algorithm of \cref{lem:decr-bundle1} has some extra log factors in the number of edges and the work/depth bound, making it less desirable for the sole purpose of maintaining a spanner. However, this extra property is useful for obtaining a $t$-bundle spanner. Previously, \cite{abraham2016fully} had exploited the same property to obtain a sequential dynamic algorithm for sparsifiers, which they called as a \textit{monotonicity property}.

To prove \cref{lem:decr-bundle1}, we use a slightly different approach for spanner computation based on \cite{miller2013parallel}. The paper proved that the following algorithm can compute the low-diameter decomposition.

\begin{algorithm}\caption{Algorithm 2 of \cite{miller2013parallel}}\label{alg:mpx13}
\begin{algorithmic}[1]
\State For each vertex $u$, pick $\delta_u$ independently from the exponential distribution Exp$(\beta)$
\State Create clusters by assigning each $v \in V$ to $u = \arg \min_{u \in V}\{dist(u, v) - \delta_u\}$, if $v = u$ we call it a center of its cluster. Let $\textsc{Cluster}(v)$ be the cluster to which $v$ belongs. Ties may be broken arbitrarily.
\State \Return $\textsc{Cluster}$
\end{algorithmic}
\end{algorithm}

\begin{lemma}[Corollary 4.5 of \cite{miller2013parallel}]\label{lem:different-pieces}
    The probability of an edge $e = (u, v)$ having $u$ and $v$ in different pieces is bounded by $O(\beta)$.
\end{lemma}

From \cref{lem:different-pieces}, we can conclude that there is a constant $\beta$ which could bound the probability of each edge $e = (u, v)$ having $u$ and $v$ in different pieces to be at most $\frac{1}{2}$. This gives the following algorithm that finds an $O(\log n)$-spanner of $O(n \log n)$: We run \cref{alg:mpx13} in $O(\log n)$ independent iterations. In each iteration, we construct a forest $H$, where each component is a spanning tree on each cluster rooted at its center. (This is the same procedure as line 5 of \cref{alg:mpvx15}). Then, the union of $H$ for each of the independent iterations is an $O(\log n)$ spanner. 

Clearly, the spanner has at most $O(n \log n)$ edges. As there is at least $\frac{1}{2}$ chance that an edge belongs to the same cluster in an iteration, for each edge, there exists an iteration where both of the edges belong to the same cluster, with high probability. Using the unique path between the vertices $u$ and $v$ in the forest $H$ of such a cluster, we can find a path of length $O(\log n)$ between $u$ and $v$ in a spanner, proving that the found spanner has $O(\log n)$ stretch. We state the above algorithm in a pseudo-code form in \cref{alg:mpx13-parallel}.

\begin{algorithm}\caption{Variant of \cite{miller2013parallel} that yields an $O(\log n)$-spanner}\label{alg:mpx13-parallel}
\begin{algorithmic}[1]
\State Set $\beta$ to be a constant where the probability of \cref{lem:different-pieces} is at most $\frac{1}{2}$.
\State $H = \{\}$
\For{$i = 0, 1, \ldots, O(\log n)$ \textit{in parallel}}
\State For each vertex $u$, pick $\delta_u$ independently from the exponential distribution Exp$(\beta)$
\State Create clusters by assigning each $v \in V$ to $u = \arg \min_{u \in V}\{dist(u, v) - \delta_u\}$, if $v = u$ we call it a center of its cluster. Let $\textsc{Cluster}(v)$ be the cluster to which $v$ belongs. Ties may be broken arbitrarily.
\State Construct a forest $H^\prime$, where each component is a spanning tree on each cluster rooted at its center.
\State Add $H^\prime$ to $H$.
\EndFor
\State \Return $H$
\end{algorithmic}
\end{algorithm}

From \cref{alg:mpx13-parallel}, it is clear that this is a parallel iteration of \cref{alg:mpvx15} limited to lines 3-5. As a result, we can use the same auxiliary directed graph used in \cref{sec:decremental-mpvx15} and apply the parallel Even-Shiloach in each of the $O(\log n)$ independent iterations. Note that we don't need to use any inter-cluster edges, nor do we need any explicit information on which cluster every vertex belongs. However, since $\delta_u$ is a real number, we still need to use the priority tag to compute the shortest path tree, as in \cref{lem:mainnear}. This concludes the description of the algorithm, and the analysis is very similar to that of \cref{lem:mainnear}.

\begin{proof}[Proof of \cref{lem:decr-bundle1}]
For each instance, we use both the directed graph $G^\prime$ and a priority tag $\textsc{Priority}$ as done in \cref{lem:mainnear}. Then, the forest $H$ is an induced subgraph of the shortest path tree over a set $V$. By definition, the probability that $\max \delta_u > 10 \frac{1}{\beta} \log n$ is at most $n^{-10}$. Hence, each shortest path tree has a depth of $O(\log n)$ with high probability. 

We first bound the total number of edges in the spanner in each instance by $O(n \log^2 n)$. As proved in \cref{lem:cluster2}, $\textsc{Priority}$ is a uniformly random permutation, independent of the integral part of $\delta_u$. For each vertex $v$ with its distance $\textsc{Dist}(v) = l$, let $U = \{u_1, u_2, \ldots, u_k\}$ be the set of vertices in $\textsc{In}(v)$ where $\textsc{Dist}(u_i) + 1 = l$. Additionally, we assume that $U$ is ordered in the following way: As the edges in the graph are removed, each $u_i$ will cease to satisfy the condition of $\textsc{Dist}(u_i) + 1 =\textsc{Dist}(v)$. We order them so that the time it ceases to satisfy the condition is nonincreasing. For example, $u_1$ will maintain the condition $\textsc{Dist}(u_i) + 1 = l$ for the longest time. Each $u_i$ will be chosen as $\textsc{Scan}(v)$, if none of $u_1, u_2, \ldots, u_{i-1}$ is placed in front of $u_i$ in the list $\textsc{In}(v)$. In our algorithm, the order we iterate $\textsc{In}(v)$ follows the permutation $\textsc{Priority}$, which is a random permutation that is independent of the order of edge removal (this follows from the oblivious adversary assumption). The probability of each $u_i$ being chosen as $\textsc{Scan}(v)$ is $\frac{1}{i}$. While the condition $\textsc{Dist}(v) = l$ holds, the expected number of $u_i$ that is ever being chosen as $\textsc{Scan}(v)$ is at most $\sum_{i=1}^{k} \frac{1}{i} \le 2 \log k \le 2 \log n$. As the depth of the shortest path is bounded by $O(\log n)$, each vertex has $O(\log^2 n)$ expected number of edges connecting to its parent in the shortest path tree, and hence the expected number of edges in the shortest path tree is at most $O(n \log^2 n)$. 

The other bounds follow from \cref{lem:mainnear} with $k = O(\log n)$, with two distinctions: First, the deletion update takes only $O(\log^2 n)$ work per instance, since we do not need to keep the cluster index. Second, since we run $O(\log n)$ instances, every work bound is multiplied as such.
\end{proof}

\subsection{Decremental $t$-bundle Spanners}
We prove the following theorem:
\decrbundle*

\begin{proof}[Proof of \cref{thm:decr-bundle2}]
    We maintain the sequence of $O(\log n)$-spanner $H_1, H_2, \ldots, H_t$ where $H_i$ is the spanner of $G \setminus (H_1 \cup \ldots \cup H_{i-1})$. Each of the spanners $H_i$ is the union of the following two components:
    \begin{itemize}
    \item A data structure $\mathcal{D}_i$, which is a decremental data structure of \cref{lem:decr-bundle1} maintaining a $O(\log n)$-spanner of $G \setminus (H_1 \cup \ldots \cup H_{i-1})$.
    \item A hash table $J_i$, a subset of edges maintained in a simple list. 
    \end{itemize}
    For the initialization, we initialize $\mathcal{D}_1, \mathcal{D}_2, \ldots, \mathcal{D}_t$ one by one, computing $H_1, H_2, \ldots, H_t$ sequentially in $O(t \log^2 n)$ depth. $J_i$ is initialized as an empty set.

    Given a deletion update in $G$, we invoke the deletion update in the data structure $\mathcal{D}_1$. As a result, $\mathcal{D}_1$ will return the set $(\delta H_{ins}, \delta H_{del})$ that is newly added or deleted into the spanner of $H_1$. We handle these sets by deleting the edges of $\delta H_{ins}$ in $\mathcal{D}_2$ and inserting the edges of $\delta H_{del}$ in $J_1$. This is possible, as any spanner will remain a spanner when extra edges in the graph are added. The procedure for $\mathcal{D}_{i = 2, 3, \ldots}$ is the same as $\mathcal{D}_1$. Given the returned set of $\delta H_{ins}$ from the data structure $\mathcal{D}_{i-1}$, we invoke the deletion update in $\mathcal{D}_i$ and put the edges of $\delta H_{del}$ in $J_i$. Finally, we check if $J_1, \ldots, J_t$ contains any edges to be deleted and delete all such edges. The set $\delta H_{ins}$ returned by $\mathcal{D}_t$, minus the set of edges given as a deletion update, represents our output $\delta H_{ins}$. The output $\delta H_{del}$ is the set of edges belonging to both the spanner $H$ and the deleted edges in the update.
    
    The algorithm's work and depth bound is $t$ times the data structure of \cref{lem:decr-bundle1}. Each set $H_i$ is the union of the spanner in $\mathcal{D}_i$ and $J_i$. $J_i$ is the union of all the edges that are returned with $\delta H_{del}$, which has its size bounded by $O(n \log^3 n)$. Hence, each spanner $H_i$ has its size bounded by $O(t n \log^3 n)$. Finally, every edge that had belonged to the $t$-bundle spanner will remain until it is removed, so each edge occurs in $\delta H_{ins}$ and $\delta H_{del}$ at most once, giving an amortized $O(1)$ size bound.
\end{proof}

\subsection{Fully Dynamic Spectral Sparsifiers}
We prove the following theorem:

\spectral*

Before proving \cref{thm:spectral}, we prove the version of \cref{thm:spectral} without incremental updates, formally stated as following:

\begin{lemma}\label{lem:spectral}
        There is a parallel batch-dynamic decremental data structure which, given an unweighted graph $G = (V, E)$ where $|V| = n, |E| = m$, maintains a $(1 \pm \epsilon)$ spectral sparsifier of at most \\ $O(n \epsilon^{-2} \log^3 m \log^6 n)$ expected number of edges. Specifically, the algorithm supports the following interfaces:

    \begin{itemize}
        \item After the initialization, the algorithm returns a set of edges forming a $(1 \pm \epsilon)$-spectral sparsifier of the given graph, which has $O(n \epsilon^{-2} \log^3 m \log^4 n)$ expected number of edges.
        \item After each edge deletion updates, the algorithm returns a pair of edge sets $(\delta H_{ins}, \delta H_{del})$, representing the set of edges that are newly inserted or deleted into the $(1 \pm \epsilon)$-spectral sparsifier. The amortized size of $|\delta H_{ins}| + |\delta H_{del}|$ is at most $O(1)$ per edge.
    \end{itemize}

    The algorithm takes:
\begin{itemize}
    \item for initialization, $O(m \epsilon^{-2} \log^3 m \log^5 n)$ expected work, and $O(\epsilon^{-2} \log^3 m \log^5 n)$ worst-case depth,
    \item for any batch of edge deletions, $O(\epsilon^{-2} \log^3 m \log^6 n)$ expected amortized work per deleted edge, and $O(\epsilon^{-2} \log^3 m \log^6 n)$ worst-case depth for the entire batch.
\end{itemize}
The above statements hold w.h.p. against an oblivious adversary.
\end{lemma}

To prove \cref{lem:spectral}, we repeat the approach of \cite{abraham2016fully} for maintaining a $(1 \pm \epsilon)$-spectral sparsifier. In \cite{abraham2016fully}, they proved that the following sequential static algorithm, named \textsc{Spectral-Sparsify}$(G, c, \epsilon)$, can compute the $(1 \pm \epsilon)$-spectral sparsifier. 

\begin{algorithm}\caption{$\textsc{Light-Spectral-Sparsify}(G, \epsilon)$ of \cite{abraham2016fully}}\label{alg:adk16_1}
\begin{algorithmic}[1]
\State $t = \Omega(\epsilon^{-2} \log^3 n)$. \Comment{For our purpose, we set $\alpha = O(\log n)$.}
\State let $B$ be a $t$-bundle spanner of $G$.
\For{$e \in G \setminus B$}
\State With probability $\frac{1}{4}$, add $e$ to $H$ with $w_H(e) \leftarrow 4 w_G(e)$
\EndFor
\State \Return $(H, B)$ \Comment{In \cite{abraham2016fully}, they return $(H\cup B, B)$ instead.}
\end{algorithmic}
\end{algorithm}

\begin{algorithm}\caption{$\textsc{Spectral-Sparsify}(G, \epsilon)$ of \cite{abraham2016fully}}\label{alg:adk16_2}
\begin{algorithmic}[1]
\State $k = \lceil \log m \rceil$ \Comment{For our purpose, we set $\rho = m$.}
\State $G_0 \leftarrow G$
\State $B_0 \leftarrow (V, \emptyset)$
\For{$i = 1, \ldots, k$}
\State $(G_i, B_i) \leftarrow \textsc{Light-Spectral-Sparsify}(G_{i-1}, \epsilon / (2k))$
\If{$G_i$ has less than $O(\log n)$ edges} \textbf{break} \Comment{Break loop.}
\EndIf
\EndFor
\State $H \leftarrow (\cup_{1\le j \le k} B_j) \cup G_k$
\State \Return $H$
\end{algorithmic}
\end{algorithm}

Although \cref{alg:adk16_1} introduces the weight $w_G(e)$ for the edges of the graph, we can ignore them in our $t$-bundle spanner computation. As long as our initial graph is unweighted, all edges given to $\textsc{Light-Spectral-Sparsify}$ as an input will have the same weight. More specifically, all edges in the graph $G_i$ will have a weight of $4^i$ if the edges of the initial graph $G_0$ have a weight of $1$. Hence, we can simply assume all edges to be unweighted and assign the weight at the very end of the computation by setting the weight in $B_j$ as $4^{j-1}$ and $G_k$ as $4^k$. 

Using these algorithms provided by \cite{abraham2016fully}, we prove \cref{lem:spectral}\shortOnly{ in the full version of the paper}.

\fullOnly{\begin{proof}[Proof of \cref{lem:spectral}]
We show the parallel implementation of \cref{alg:adk16_2} under edge deletion updates that satisfy the bounds in the statement. For each graph $G_0, G_1, \ldots, G_k$, we maintain the decremental $t$-bundle spanner $B_i$ using the data structure $\mathcal{B}_i$ of \cref{thm:decr-bundle2}. In the initialization stage, we compute each of the $G_0, G_1, \ldots, G_k$ by initializing $\mathcal{B}_i$ over $G_i$, and then sampling the edges of $G_i \setminus B_i$ with probability $\frac{1}{4}$ to populate $G_{i+1}$. 

Given a deletion update on $G_i$, we invoke the decremental update to $\mathcal{B}_i$. As a result, $\mathcal{B}_i$ will return a pair of set of edges $(\delta H_{ins}, \delta H_{del})$. $\delta H_{del}$ can be ignored since they are a subset of our deletion update and are not inserted in $G_{i+1}, G_{i+2}, \ldots$. $\delta H_{ins}$ contains the set of edges newly inserted to the $t$-bundles of $B_i$ and hence should be removed from the $G_{i+1}$. Since every edge of $G_{i+1}$ is sampled uniformly and independently from $G_i$, we can add the edges from $\delta H_{ins}$ to our current set of deletion update, filter only the edges that are sampled in $G_{i+1}$, and invoke the decremental update in the data structure $\mathcal{B}_{i+1}$. If the decremental update reduces the number of edges in $G_i$ below $O(\log n)$, we destroy the data structure and reduce the $k$ accordingly. To compute $\delta H_{ins}$ and $\delta H_{del}$, we take the union of all $\delta H_{ins}$ and $\delta H_{del}$ from the data structure $\mathcal{B}_1, \ldots, \mathcal{B}_k$ (with multiplicity, as in a multiset) and remove the intersection.

The work and depth bound now follow from \cref{thm:decr-bundle2}, by setting $t = O(\epsilon^{-2} \log^2 m \log^3 n)$, and multiplying all work and depth bound by $\log m$. Finally, in our algorithm, every edge that had belonged to the $(1 \pm \epsilon)$ spectral sparsifier will remain until it is removed, so each edge occurs in $\delta H_{ins}$ and $\delta H_{del}$ at most once, giving an amortized $O(1)$ size bound. 
\end{proof}
}
We turn our decremental data structure into a fully-dynamic one using the same method we used in \cref{sec:full-mpvx15}. In the full-version paper of \cite{abraham2016fully}, they prove the following property:

\begin{lemma}[Lemma 4.18 of \cite{abraham2016fully}]\label{lem:decomposable}
    Let $G = (V, E)$ be an undirected weighted graph, let $E_1, \ldots, E_k$ be a partition of edge set $E$, and let, for every $1 \le i \le k$, $H_i$ be a $(1 \pm \epsilon)$-spectral sparsifier of $G_i = (V, E_i)$. Then $H = \cup_{i=1}^k H_i$ is a $(1 \pm \epsilon)$-spectral sparsifier of $G$.
\end{lemma}

Using \cref{lem:spectral} and \cref{lem:decomposable}, we can repeat the proof of \cref{thm:mainnear} to obtain \cref{thm:spectral}.

\begin{proof}[Proof of \cref{thm:spectral}]
The fully dynamic algorithm maintains the partition of edges $E = E_0 \cup E_1 \cup \ldots \cup E_b$, where $b \leq O(\log m)$. Additionally, the algorithm maintains a global hash table $\textsc{Index}(e)$ which, given an edge $e \in E$, maintains the index $i$ where $e \in E_i$. Let $l_0$ be the smallest integer such that $2^{l_0} \geq n$. Each partition $E_i$ satisfies the following invariant:

\paragraph{Invariant B2} $|E_i| \leq 2^{i + l_0}$.

Then, we proceed everything identically as in \cref{thm:mainnear}, where the invariant B1 is replaced with invariant B2, and the decremental spanner algorithm is replaced with the decremental $(1 \pm \epsilon)$ spectral sparsifier algorithm of \cref{lem:spectral}. 

The work and depth bound for deletion and initialization remain the same with \cref{lem:spectral}. The work and depth bound for insertion comes from the initialization bound of \cref{lem:spectral}, with the same depth and $O(\log m)$ times the work for initialization. As every edge can belong to at most $O(\log m)$ different decremental instance, the amortized total size of $|\delta H_{ins}| + |\delta H_{del}|$ is at most $O(\log m)$. 
\end{proof}

\bibliographystyle{alpha} 
\bibliography{library}

\end{document}